\documentclass[prd,aps,twocolumn,preprintnumbers, showpacs, nofootinbib,superscriptaddress,notitlepage]{revtex4-1}
\usepackage{amsmath,graphicx,epsfig,amssymb}
\usepackage{inputenc}
\usepackage{tabularx}
\usepackage[usenames]{color}
\usepackage{ulem} 
\usepackage{bigstrut}
\usepackage{slashed}
\usepackage{multirow}
\usepackage{subfigure}
\usepackage{dsfont}
\usepackage{xcolor}

\allowdisplaybreaks

\begin{document}

\title{Light Cone Distribution Amplitude for the $\Lambda$ Baryon from Lattice QCD}
\collaboration{\bf{Lattice Parton Collaboration ($\rm {\bf LPC}$)}}

\author{\includegraphics[scale=0.10]{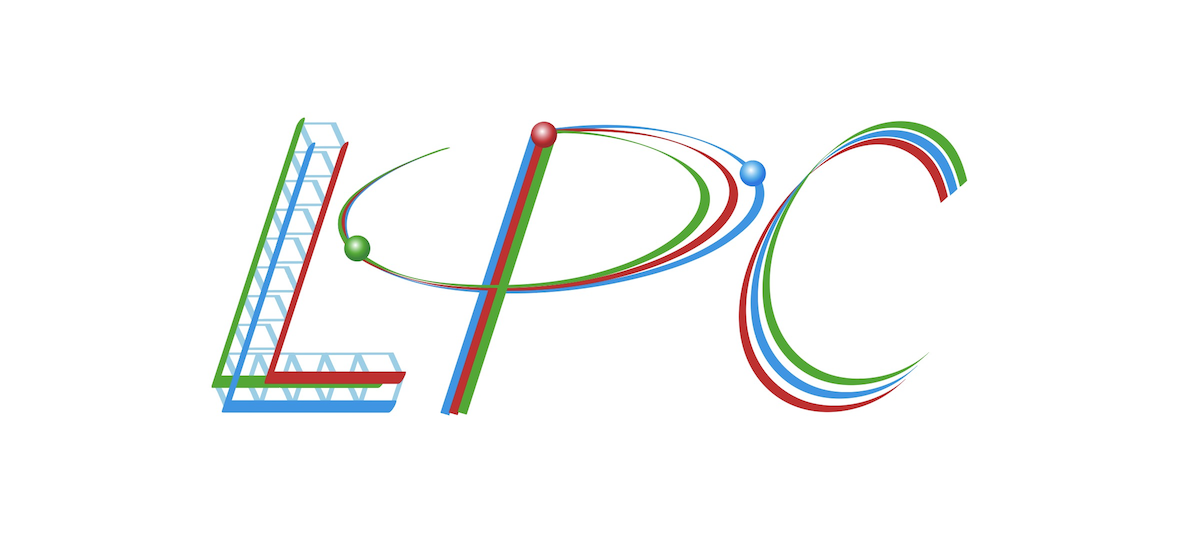}\\Min-Huan Chu}
\affiliation{INPAC, Key Laboratory for Particle Astrophysics and Cosmology (MOE),  Shanghai Key Laboratory for Particle Physics and Cosmology, School of Physics and Astronomy, Shanghai Jiao Tong University, Shanghai 200240, China}
\affiliation{Yang Yuanqing Scientific Computering Center, 
Tsung-Dao Lee Institute, Shanghai Jiao Tong University, Shanghai 200240, China}
\affiliation{Faculty of Physics, Adam Mickiewicz University, ul. Uniwersytetu Pozna\'nskiego 2, 61-614 Pozna\'n, Poland}

\author{Haoyang Bai}
\affiliation{Institute of High Energy Physics, CAS, Beijing 100049, China}
\affiliation{School of Physics, University of Chinese Academy of Sciences, Beijing 100049, China}

\author{Jun Hua}
\email{Corresponding author: junhua@scnu.edu.cn}
\affiliation{Key Laboratory of Atomic and Subatomic Structure and Quantum Control (MOE), 
Guangdong Basic Research Center of Excellence for Structure and Fundamental Interactions of Matter, 
Institute of Quantum Matter, South China Normal University, Guangzhou 510006, China}
\affiliation{Guangdong-Hong Kong Joint Laboratory of Quantum Matter, 
Guangdong Provincial Key Laboratory of Nuclear Science, Southern Nuclear Science Computing Center, 
South China Normal University, Guangzhou 510006, China}

\author{Jian Liang}
\affiliation{Key Laboratory of Atomic and Subatomic Structure and Quantum Control (MOE), Guangdong Basic Research Center of Excellence for Structure and Fundamental Interactions of Matter, Institute of Quantum Matter, South China Normal University, Guangzhou 510006, China}
\affiliation{Guangdong-Hong Kong Joint Laboratory of Quantum Matter, Guangdong Provincial Key Laboratory of Nuclear Science, Southern Nuclear Science Computing Center, South China Normal University, Guangzhou 510006, China}

\author{Xiangdong Ji}
\affiliation{Department of Physics, University of Maryland, College 
Park, MD 20742, USA}

\author{Andreas Sch\"afer}
\affiliation{Institut f\"ur Theoretische Physik, Universit\"at Regensburg, D-93040 Regensburg, Germany}

\author{Yushan Su}
\affiliation{Department of Physics, University of Maryland, College Park, MD 20742, USA}

\author{Wei Wang}
\email{Corresponding author: wei.wang@sjtu.edu.cn}
\affiliation{INPAC, Key Laboratory for Particle Astrophysics and Cosmology (MOE),  Shanghai Key Laboratory for Particle Physics and Cosmology, School of Physics and Astronomy, Shanghai Jiao Tong University, Shanghai 200240, China}
\affiliation{Southern Center for Nuclear-Science Theory (SCNT), Institute of Modern Physics, Chinese Academy of Sciences, Huizhou 516000, Guangdong Province, China}

\author{Yi-Bo Yang}
\affiliation{CAS Key Laboratory of Theoretical Physics, Institute of Theoretical Physics, Chinese Academy of Sciences, Beijing 100190, China}
\affiliation{School of Fundamental Physics and Mathematical Sciences, Hangzhou Institute for Advanced Study, UCAS, Hangzhou 310024, China}
\affiliation{International Centre for Theoretical Physics Asia-Pacific, Beijing/Hangzhou, China}
\affiliation{School of Physical Sciences, University of Chinese Academy of Sciences,
Beijing 100049, China}

\author{Jun Zeng}
\affiliation{INPAC, Key Laboratory for Particle Astrophysics and Cosmology (MOE),  Shanghai Key Laboratory for Particle Physics and Cosmology, School of Physics and Astronomy, Shanghai Jiao Tong University, Shanghai 200240, China}

\author{Jian-Hui Zhang}
\affiliation{School of Science and Engineering, The Chinese University of Hong Kong, Shenzhen 518172, China}

\author{Qi-An Zhang}
\affiliation{School of Physics, Beihang University, Beijing 102206, China}

\begin{abstract}
We calculate the leading-twist light-cone distribution amplitudes of the light $\Lambda$ baryon using lattice methods within the framework of large momentum effective theory. Our numerical computations are conducted employing $N_f=2+1$ stout smeared clover fermions and a Symanzik gauge action on a lattice with spacing $a=0.077\;\rm{fm}$, and a pion mass of 303 MeV. To approach the large momentum regime, we simulate the equal-time correlations with the hadron momentum $P^z = \{2.52, 3.02, 3.52\}$ GeV. By investigating the potential analytic characteristics of the baryon quasi-distribution amplitude in coordinate space, we validate these findings through our lattice calculations. After renormalization and extrapolation,  we present results for the three-dimensional distribution of momentum fractions for the two light quarks.   Based on these findings the paper briefly discusses the phenomenological impact on weak decays of $\Lambda_b$, and outlines potential systematic uncertainties  that can be improved in the future.  This work  lays the theoretical foundation for accessing baryon LCDAs from lattice QCD.
\end{abstract}

\maketitle

\section{ Introduction}
The light-cone distribution amplitudes (LCDAs)  are  probability  amplitudess for the longitudinal momentum fractions of partons in the leading Fock states of hadrons~\cite{Lepage:1980fj,Chernyak:1983ej}. Within Quantum Chromodynamics (QCD), the non-perturbative LCDAs play a pivotal role for the description of exclusive processes (e.g. heavy baryon decays) and allow, e.g., to calculate formfactors. They provide information on hadron structure which is complementary to that of parton distribution functions.
 Thus, lattice calculations of LCDAs support not only experimental programs at facilities like the Large Hadron Collider (LHC) and Jefferson Lab, but by accurately determining the shape and properties of baryon LCDAs and by predicting various observables, e.g., of heavy baryon decays, one can also test theoretical models, ultimately deepening our understanding of the strong interactions that govern the behavior of baryons. An example of this phenomenon might be the predicted occurrence of CP violation in heavy baryon decays, with some suggestive clues emerging from recent observations by the LHCb collaboration~\cite{LHCb:2016yco}.  Generally speaking, the advent of high-luminosity accelerators has transformed the investigation of exclusive reactions, and thus also of LCDAs, into a highly fertile research field.

In the past decades,  significant progress has been made in characterizing the LCDAs of light mesons,  including both a few lowest moments \cite{Bali:2017ude, RQCD:2019osh} and the complete $x$-dependent distribution \cite{Zhang:2017bzy,Chen:2017gck,Zhang:2020gaj,Gao:2022vyh,Holligan:2023rex,Hua:2020gnw, LatticeParton:2022zqc,Baker:2024zcd,Cloet:2024vbv}.  
Conversely, there has been limited advancement in determining baryon LCDAs due to substantial obstacles.  The three valence quarks in a baryon interact with each other through strong forces, leading to a rich and intricate structure that is difficult to describe using simple models.  The earliest estimates of the first and second  moments of the baryon LCDAs were made over 30 years ago based on the QCD sum rule, known as the Chernyak-Ogloblin-Zhitnitsky (COZ) model \cite{Chernyak:1987nu}. This result provided a reasonable description of the experimental data at the time,  but gave a distribution amplitude that deviated significantly from the asymptotic behavior.  Meanwhile, advancements in perturbation theory have led to widespread consensus that the analysis of form factors in exclusive processes with substantial momentum transfer~\cite{Radyushkin:1990te, Bolz:1996sw} does not indicate a notable deviation of the baryon’s light-cone distribution amplitudes from their asymptotic form \cite{Braun:2001tj, Anikin:2013aka}. Recently, new methods such as the On-shell wavefunction method, have proposed new strategies, but the results are still model-dependent~\cite{Bell:2013tfa}.

Lattice QCD  provides first-principle tools to access baryon LCDAs, offering insights into the internal structure of baryons and paving the way for a more comprehensive understanding of baryonic systems.  Notably, the lowest moments of the leading twist baryon LCDAs have been computed using operator product expansion (OPE) on the lattice \cite{Bali:2015ykx, RQCD:2019hps}. But unfortunately, in intricate exclusive processes like heavy baryon decay, existing phenomenological analyses do not definitively establish the dominance of the leading-order  contributions, and, therefore, requires more information than just a few moments. In Ref.~\cite{Chen:2024fhj,Huang:2024ugd}, systematic explorations of the next-to-leading order contributions to nucleon form factors have been undertaken, in which the NLO corrections are found to be significant   
 and the nonperturbative evaluation of the entire shape of the baryon LCDAs is necessary.

In recent years, the development of  large momentum effective theory (LaMET) has presented a promising new approach to tackle the lightcone distributions. More explicitly,  LaMET offers a framework for computing quasi-distributions, which are equal-time correlation functions in large-momentum hadron states displaying the same infrared characteristics  as light-cone distributions. By establishing a connection between these quasi-distributions and  light-cone distributions through a matching scheme, LaMET provides a pathway for deriving baryon LCDAs from lattice QCD calculations. Since LaMET was proposed it has a wide range of applications, including calculations of quark distribution functions~\cite{Xiong:2013bka,Lin:2014zya,Alexandrou:2015rja,Chen:2016utp,Alexandrou:2016jqi,Alexandrou:2018pbm,Chen:2018xof,Lin:2018pvv,LatticeParton:2018gjr,Alexandrou:2018eet,Liu:2018hxv,Chen:2018fwa,Izubuchi:2018srq,Izubuchi:2019lyk,Shugert:2020tgq,Chai:2020nxw,Lin:2020ssv,Fan:2020nzz,Gao:2021hxl,Gao:2021dbh,Gao:2022iex,Su:2022fiu,LatticeParton:2022xsd,Gao:2022uhg,Chou:2022drv,Gao:2023lny,Gao:2023ktu,Chen:2024rgi,Holligan:2024umc,Holligan:2024wpv}, gluon distribution functions~\cite{Wang:2017qyg,Wang:2017eel,Fan:2018dxu,Wang:2019tgg,Good:2024iur}, generalized parton distributions~\cite{Chen:2019lcm,Alexandrou:2019dax,Lin:2020rxa,Alexandrou:2020zbe,Lin:2021brq,Scapellato:2022mai,Bhattacharya:2022aob,Bhattacharya:2023nmv,Bhattacharya:2023jsc,Lin:2023gxz,Holligan:2023jqh,Ding:2024hkz}, lightcone distribution amplitudes of a light hadron~\cite{Zhang:2017bzy,Chen:2017gck,Zhang:2020gaj,Hua:2020gnw, LatticeParton:2022zqc,Gao:2022vyh,Holligan:2023rex,Liu:2018tox,Deng:2023csv,Han:2023xbl,Han:2023hgy,Han:2024ucv,Baker:2024zcd,Cloet:2024vbv} and  a heavy meson~\cite{Wang:2019msf,Zhao:2020bsx,Xu:2022krn,Xu:2022guw,Hu:2023bba,Hu:2024ebp,Han:2024min,Han:2024cht,Deng:2024dkd,Han:2024yun},  transverse-momentum-dependent distributions~\cite{Ji:2014hxa,Shanahan:2019zcq,Shanahan:2020zxr,Zhang:2020dbb,Ji:2021znw,LatticePartonLPC:2022eev,Liu:2022nnk,Zhang:2022xuw,Deng:2022gzi,Zhu:2022bja,LatticePartonCollaborationLPC:2022myp,Rodini:2022wic,Shu:2023cot,Chu:2023jia,delRio:2023pse,LatticePartonLPC:2023pdv,LatticeParton:2023xdl,Alexandrou:2023ucc,Avkhadiev:2023poz,Zhao:2023ptv,Avkhadiev:2024mgd,Bollweg:2024zet,Spanoudes:2024kpb}, and double parton distribution functions~\cite{Zhang:2023wea,Jaarsma:2023woo}. In contrast with the calculation of lowest  moments using the OPE, LaMET enables the extraction of complete $x$-dependent distribution amplitudes from first principles through lattice computations. These studies of nucleons parton distribution functions  and light meson light-cone distribution amplitudes   have  confirmed the validity of LaMET. Additionally, research on three-dimensional distribution amplitudes  further underscores the significant potential of LaMET.

This paper  pioneers lattice QCD computations of baryon LCDAs within the LaMET framework,  elucidates the theoretical underpinnings, andcomputational strategies, and  presents some numerical results which serve as proof of principle that such calculations are feasible. Future work will improve the precision and enlarge the scope of such calculations, for example by analyzing several ensembles and performing a controlled extrapolation to the physical point, and will thus bridge the gap to a comparison with experimental data.



The subsequent sections of this paper are structured as follows. In Sec.~\ref{sec:theroetical_framework}, we give the theoretical framework including  the definitions of LCDAs and quasi DAs, renormalizations, symmetry properties of quasi DAs, and extrapolations of the results. Section III is dedicated to presenting the numerical outcome of  the lattice QCD simulation. Based on these findings the paper briefly discusses the phenomenological impact on weak decays of $\Lambda_b$, and identifies potential systematic uncertainties that then will be reduced in the future. The final section includes the summary and prospects. Some calculation details are provided in the appendix.

\section{Theoretical framework}
\label{sec:theroetical_framework}

\subsection{The LCDA and quasi-DA for the $\Lambda$ baryon}

\begin{figure}
\centering
\includegraphics[scale=0.1]{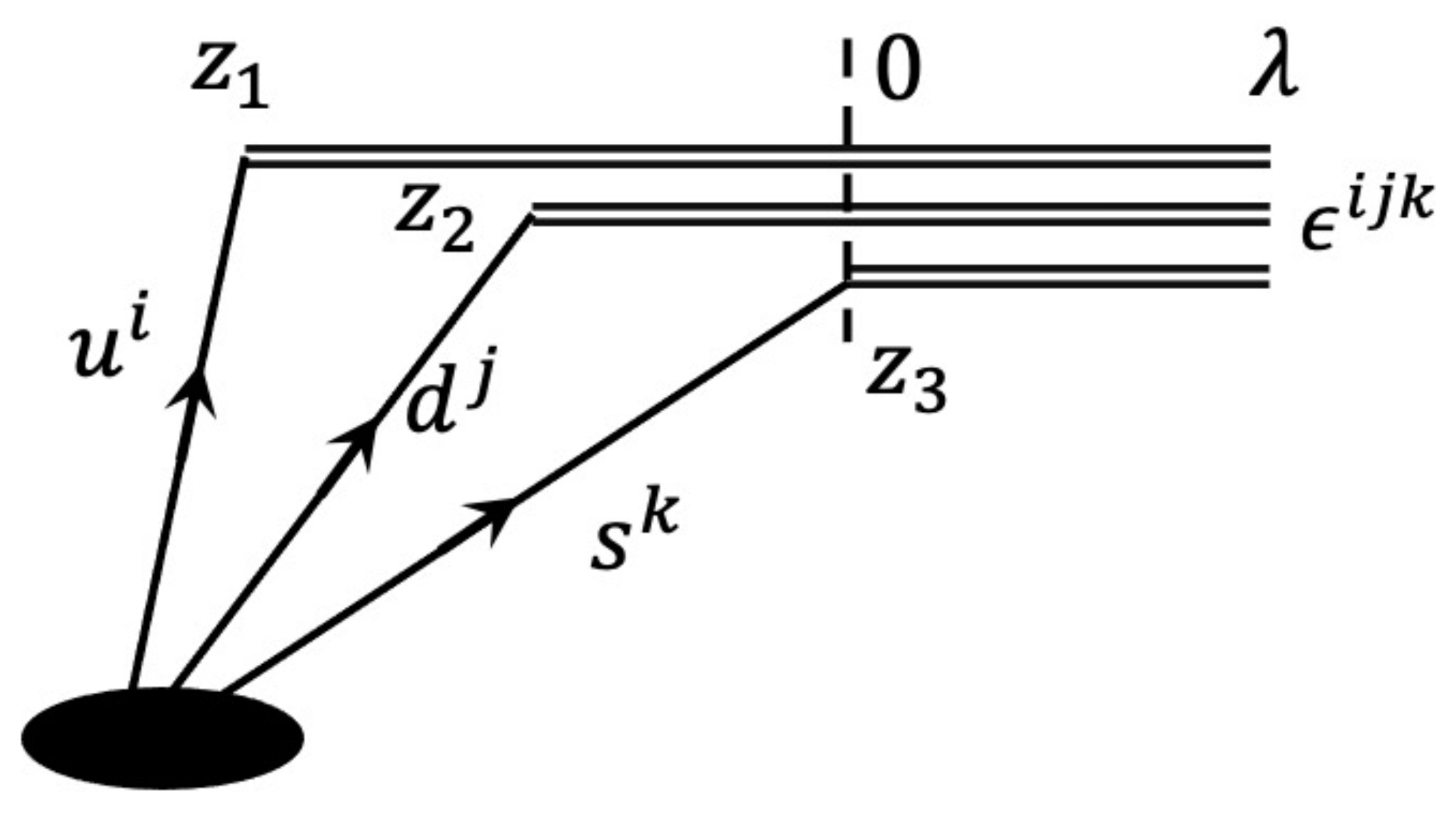}
\caption{The structure for the  $\Lambda$ LCDAs, with three Wilson lines connected to the three quarks. $z_3$ can be set to 0 for simplification.}
\label{fig:baryonplot}
\end{figure}

The leading LCDAs for the $\Lambda$ baryon is given by the non-local hadron-to-vacuum matrix elements at light-like separations:
\begin{align}
&H(z_1,z_2,z_3)_{\alpha \beta \gamma} = \varepsilon^{i j k} \langle 0|  (W^{i i^{\prime}}\left(z_0n, z_1n\right)) u_\alpha^{i^{\prime}}\left(z_1n\right)   \nonumber \\ &\;\;\; (W^{ j j^{\prime}}\left(z_0n, z_2n\right)) d_\beta^{j^{\prime}}\left(z_2 n\right) \nonumber \\
&\;\;\;(W^{k k^{\prime} }\left(z_0n, z_3n\right)) s_\gamma^{k^{\prime}}\left(z_3n\right) |\Lambda(P,\lambda)\rangle,
\label{eq:baryon_matrix}
\end{align}
where $|\Lambda(P,\lambda)\rangle$ represents the $\Lambda$ state with momentum $P$ and helicity $\lambda$. $W_{ij}(a,b)$ are light-like Wilson lines to preserve gauge invariance, and $z_0$ is a reference position on the Wilson lines.  The subscripts $\alpha,\beta,\gamma$ refer to spin indices and the superscripts $i,j,k$ refer to color indices.  $n$ is an arbitrary light-like vector $n^2 = 0$. An illustration of this structure is given in  Fig.~\ref{fig:baryonplot}.

A most general decomposition of the matrix element in Eq.~(\ref{eq:baryon_matrix}) involves 24 invariant functions~\cite{Braun:2000kw}, but only three of them are of leading twist (twist-3):
\begin{eqnarray}
&& H(z_1,z_2,z_3)_{\alpha \beta \gamma} =\frac{1}{4}f_V\big[(\slashed{P} C)_{\alpha \beta}\left(\gamma_5 u_{\Lambda}\right)_\gamma V\left(z_i P \cdot n\right) \nonumber\\
&&\;\;\; +\left(\slashed{P} \gamma_5 C\right)_{\alpha \beta}\left(u_{\Lambda}\right)_\gamma A\left(z_i P \cdot n\right)\big] \nonumber\\
&&\;\;\; +\frac{1}{4}f_T\big[\left(i \sigma_{\mu \nu} P^\nu C\right)_{\alpha \beta}\left(\gamma_\mu \gamma_5 u_{\Lambda}\right)_\gamma T\left(z_i P \cdot n\right)\big].
\label{eq:leading_twist}
\end{eqnarray}

By inserting a suitable Dirac matrix $\Gamma$ the different leading twist functions can be projected out:
\begin{align}
&\epsilon_{ijk}\left\langle 0\left|u^{i, \rm T}\left(z_1 n\right) \Gamma d^j\left(z_2 n\right) s^k(z_3n)\right| \Lambda\right\rangle\nonumber\\
&=\Phi\left(z_1,z_2,\mu\right)P^+
f_{\Lambda}u_{\Lambda}(P),
\label{eq:lcda_co}
\end{align}
where $\mu$ accounts for the renormalization scale. To simplify the notation, we can set $z_3$ to be zero as shown in Fig.~\ref{fig:baryonplot}. Therefore the normalized LCDA as a function of momentum fractions in momentum space can be defined as:
\begin{align}
\phi(x_1,x_2,\mu)=&\int\frac{P^+dz_1}{2\pi}\int\frac{P^+dz_2}{2\pi}~e^{-i(x_1z_1+x_2z_2)P^z}\nonumber \\ 
&\times \Phi(z_1,z_2,\mu),
\label{eq:lcda_mom}
\end{align}
where $x_{1,2}$ denote the longitudinal momentum fractions carried by two light quarks.  Thereby the momentum fraction of the strange quark  is $1-x_1-x_2$. In this work, we only consider  the leading twist component $A(z_iP\cdot n)$ which can be projected out using   $\Gamma=C \gamma_5 \slashed{n}$.  
This is a severe limitation, as \cite{RQCD:2019hps} suggests (based, however, solely on the zeroth and first moment) that $A$, $V$ and $T$ are of comparable size. Therefore, this study is primarily a proof of principle that such calculations are feasible in LaMET. $T$ and $V$ will be calculated in future work.


In LaMET the LCDAs can be accessed via the   lattice simulation of an equal time matrix element for $\Lambda$ defined as
\begin{align}
&\tilde{\Phi}_A\left(z_1,z_2,\mu\right) P^z
f_{\Lambda} u_{\Lambda}(P^z) = \nonumber\\
&\epsilon_{i j k}\left\langle 0\left|u^{i, \rm T}\left(z_1 n_z\right) \tilde{\Gamma} d^j\left(z_2 n_z\right) s^k(z_3n_z)\right| \Lambda(P^z)\right\rangle.
\label{eq:matrix_element}
\end{align}
This function is similar to the light-cone distribution in Eq.~(\ref{eq:lcda_co}), while $n_z=(0,0,0,1)$ is the unit vector along the $z$-direction in Euclidean space, the same direction as the momentum $P^z$. The Dirac matrix $\tilde{\Gamma}=C \gamma^5 \gamma^{\nu}$ can be chosen as $\nu=t$ or $\nu=z$, both of which could approach the leading-twist $C\gamma_5\gamma^+$ on the light-cone in the large momentum limit. In this work, we choose $C\gamma^5\gamma^t$, and   call $\tilde{\Phi}_A(z_1,z_2)$ quasi-DA in coordinate space.  The other two LCDAs will be left for future study. 

The quasi-distribution amplitude  in momentum space can be obtained from the distribution in coordinate space by a Fourier transformation:
\begin{align}
\tilde{\phi}_A(x_1,x_2,\mu)=&\int\frac{P^zdz_1}{2\pi}\int\frac{ P^zdz_2}{2\pi}~e^{-i(x_1z_1+x_2z_2)P^z} \nonumber\\
&\times\tilde{\Phi}_A(z_1,z_2,\mu).
\label{eq:quasi_co}
\end{align}

The  correlation function that can be used to extract the quasi DAs are defined as: 
\begin{align}
C_2(z_1,z_2;t,\vec{P})&=\int d^3xe^{-i\vec{P}\vec{x}}\langle0|\hat{O}_{\gamma}(\vec{x},t;z_1,z_2)\nonumber\\
&\times{\bar{\hat O}}_{\gamma'}(0,0;0,0)T^{\gamma\gamma'}|0\rangle,
\label{eq:2pt_definition}
\end{align}
where $T$ denotes the projection operator extracting matrix components for correlation functions and is chosen as $(I+\gamma^t)/2$ for $\Lambda$ with $I(J^P)=0(\frac{1}{2}^+)$. We denote the baryon operator as $\hat{O}_{\gamma}(\vec{x},t;z_1,z_2)$. It consists of three quark fields and three Wilson lines:
\begin{align}
\hat{O}_{\gamma}(\vec{x},t;z_1,z_2)&=\epsilon^{ijk}W^{ii'}(z_0,\vec{x}+z_1n_z)u^{i'}_{\alpha}(\vec{x}+z_1n_z,t)\nonumber\\
&\times \tilde\Gamma_{\alpha\beta}W^{jj'}(z_0,\vec{x}+z_2n_z,t) d^{j'}_{\beta}(\vec{x}+z_2n_z)\nonumber\\
&\times W^{kk'}(z_0,\vec{x})s^{k'}_{\gamma}(\vec{x},t),
\label{eq:2pt_cor}
\end{align}
where   $W(z_0,\vec{x})$    denotes the Wilson line in coordinate space.  We have  chosen  $z_3$  to be zero for simplification.

The reference position $z_0$ of Wilson lines in $\hat{O}_{\gamma}(\vec{x},t;z_1,z_2)$ is arbitrary, and for simplicity, we set it to the same value as that of the strange quark. Therefore, the operator can be expressed as:
\begin{align}
\hat{O}_{\gamma}(\vec{x},t;z_1,z_2)&=\epsilon^{ijk}W^{ii'}(\vec{x},\vec{x}+z_1n_z)u^{i'}_{\alpha}(\vec{x}+z_1n_z,t)\nonumber\\
&\times \tilde\Gamma_{\alpha\beta}W^{jj'}(\vec{x},\vec{x}+z_2n_z)d^{j'}_{\beta}(\vec{x}+z_2n_z,t)\nonumber\\
&\times s^k_{\gamma}(\vec{x},t). 
\end{align}

\subsection{Renormalization}
The non-local operator in correlation functions  includes Wilson lines which contain both linear and logarithmic ultraviolet (UV) divergences, and a non-perturbative renormalization scheme to remove these divergences has to be adopted.
Over the last few years, several renormalization schemes have been proposed and applied to try to eliminate these divergences, including Wilson-line renormalization \cite{Zhang:2017bzy,Zhang:2017zfe}, RI/MOM renormalization \cite{Martinelli:1994ty,Zhang:2020gaj}, hybrid renormalization\cite{Ji:2020brr,Hua:2020gnw} and self renormalization \cite{LatticePartonLPC:2021gpi,LatticeParton:2022zqc}.

The RI/MOM renormalization scheme has been extensively used in the renormalization of local operators, but its application to non-local operators is a topic of ongoing debate. Recent studies~\cite{Zhang:2024omt} indicate that the RI/MOM scheme is highly reliant on the precision of gauge fixing. When gauge fixing is sufficiently accurate, the results in the RI/MOM scheme agree with perturbative results at short distances. However, for large distqnces the RI/MOM scheme still introduces artificial infrared effects.

The hybrid renormalization approach~\cite{Ji:2020brr} introduced a novel strategy by implementing different renormalization schemes in distinct coordinate regions, offering a promising framework for addressing the renormalization of non-local operators. The concept of self-renormalization~\cite{LatticePartonLPC:2021gpi} is based on the foundations of hybrid renormalization, utilizing the self-ratio approach at short distances, which involves the comparison of large-momentum matrix elements to zero-momentum matrix elements. This method involves extracting linear divergences and other parameters from zero-momentum matrix elements at various lattice spacings through matching with perturbative results at short distances. Subsequently, mass renormalization can be applied in the long-distance regime. As a result, self-renormalization represents an advancement in Wilson-line renormalization techniques.

The hybrid scheme based on self-renormalization is currently the mainstream scheme for studying LCDAs with LaMET \cite{LatticeParton:2022zqc,Baker:2024zcd,Cloet:2024vbv}. Some details of hybrid renormalization for light baryons have already been discussed analytically in \cite{Han:2023xbl,Han:2024ucv}, however the numerical application is quite involved. Firstly, to extract the linear divergence, multiple zero-momentum matrix elements at different lattice spacings need to be calculated. More importantly, self-renormalization requires matching with perturbative results at short-distance to extract the renormalization factors. According to \cite{Han:2023xbl}, the perturbative zero momentum matrix element of $\Lambda$ is:
\begin{align}
 \hat{M}_p&\left(z_1, z_2, 0,0, \mu\right)=1+\frac{\alpha_s C_F}{2 \pi}\bigg[\frac{7}{8} \ln \left(\frac{z_1^2 \mu^2 e^{2 \gamma_E}}{4}\right) \nonumber \\
+ & \frac{7}{8} \ln \left(\frac{z_2^2 \mu^2 e^{2 \gamma_E}}{4}\right)+\frac{3}{4} \ln \left(\frac{\left(z_1-z_2\right)^2 \mu^2 e^{2 \gamma_E}}{4}\right)+4 \bigg].
\label{eq:phi_co_per}
\end{align}
Divergences exist at $z_1=0, z_2=0$ and $z_1=z_2$. Matching the divergent points in numerical calculations with those in analytical calculations is complicated. We need to match with perturbative results in the short-distance region $ (<0.3 \; \rm{fm})$ while avoiding these divergences. This implies that very small lattice spacings and highly precise calculations are required to achieve self-renormalization for a baryon. Therefore, in this work, we use the ratio of large momentum to zero-momentum matrix elements for renormalization, ignoring the systematic errors introduced by spurious infrared physics at long distances. The renormalized quasi-DA can be expressed as: 
\begin{align}
\tilde{\Phi}_A(z_1,z_2,\mu,P^z\ne0)=\frac{\tilde{\Phi}_A^0(z_1,z_2,P^z\ne0)}{\tilde{\Phi}_A^0(z_1,z_2,P^z=0)}.
\end{align}

\subsection{Analytic properties of quasi DAs}

Before initiating numerical simulations,  we can leverage certain analytic properties to streamline our calculations.

\begin{figure}
\centering
\includegraphics[scale=0.3]{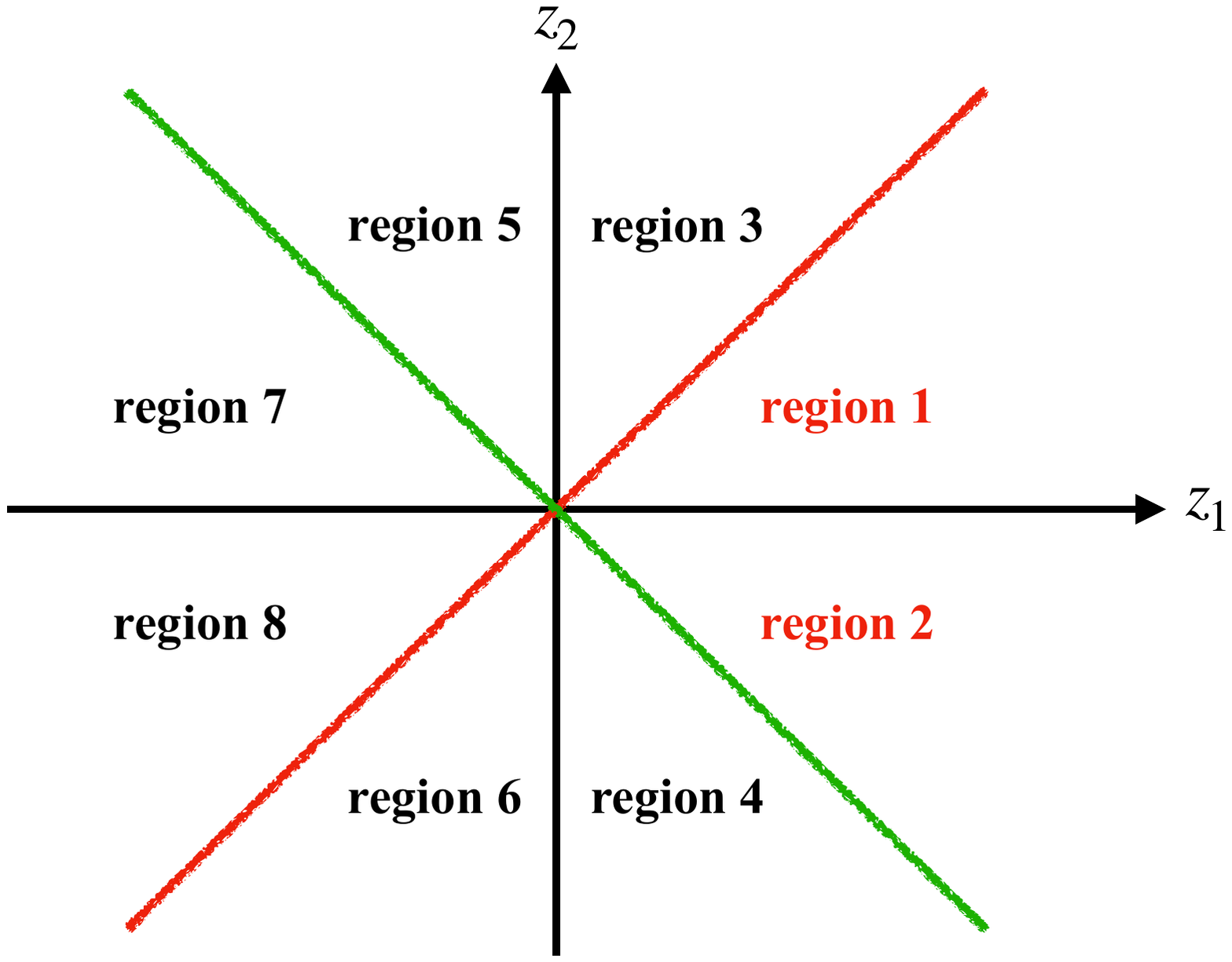}
\caption{Eight regions in $z_1-z_2$ plane.}
\label{fig:sym_region}
\end{figure}

To discuss the analytical properties of $\tilde{\Phi}(z_1,z_2)$ we distinguish eight regions, as shown in Fig.~\ref{fig:sym_region} and use two symmetries to simplify the calculation.
\begin{itemize}
\item The isospin symmetry, results from the fact that in lattice simulations we do not distinguish u and d quarks of our clover fermion ensembles. As $z_1$ and $z_2$ are the coordinate of the u and d quarks, we have the equality  
\begin{align}
\tilde{\Phi}(z_1,z_2)=\tilde{\Phi}(z_2,z_1).
\label{eq:sys_1}
\end{align}
\item Another constraints arises from the fact that the LCDAs and the quasi DAs in momentum space are real : 
\begin{align}
\tilde{\phi}(z_1,z_2,\mu)=& \int^1_0 dx_1 \int^1_0 dx_2 e^{i(x_1z_1P^z+x_2z_2P^z)} \nonumber\\
&\times\tilde{\Phi}(x_1,x_2,\mu),
\label{eq:quasi_mom}
\end{align}
the real part of the quasi-DA in coordinate space is symmetric, while the imaginary part is antisymmetric. The $\lambda_{1,2}$ in this equation denotes the light front (LF) distance $\lambda_{1,2} = z_{1,2}P^z$. Then $\tilde{\Phi}(z_1,z_2)$ satisfies:
\begin{align}
\tilde{\Phi}(z_1,z_2)=\tilde{\Phi}^*(-z_1,-z_2).
\label{eq:sys_2}
\end{align}
\end{itemize}
Therefore, the symmetry of the eight regions in Fig~.\ref{fig:extra_area} can be summarized as follows:
\begin{align}
&\tilde{\Phi}_{1}(z_1,z_2) = \tilde{\Phi}_{3}(z_2,z_1) = \tilde{\Phi}^*_{6}(-z_2, -z_1) = \tilde{\Phi}^*_{8}(-z_1,-z_2), \nonumber \\
&\tilde{\Phi}_{2}(z_1,z_2) = \tilde{\Phi}^*_{4}(-z_2,-z_1) = \tilde{\Phi}_{5}(z_2,z_1) = \tilde{\Phi}^*_{7}(-z_1,-z_2).
\label{eq:sym_prop}
\end{align}

Based on these symmetries, only regions 1 and 2 are independent. Given that the propagators for u and d quarks are identical in our lattice simulation, the first symmetry holds true for each ensemble. The second symmetry stems from the physical constraints imposed on the quasi-distribution amplitude in momentum space, resulting in a statistical symmetry observed in the lattice data.  Consequently, data from region 4 and region 6 can also be generated on the lattice, and subsequently merged with data from region 1 and region 2 to enhance statistical significance. Furthermore, as a result of these symmetries along the diagonal line where $z_1 = -z_2$,  represented by the green line in Fig. \ref{fig:sym_region}, it holds that $\mathrm{Im}[\tilde{\Phi}(z_1,z_2)] \equiv 0$.

\subsection{Extrapolation}
\label{sec:frame_extra}
\label{sec:extra}

\begin{figure}
\centering
\includegraphics[scale=0.12]{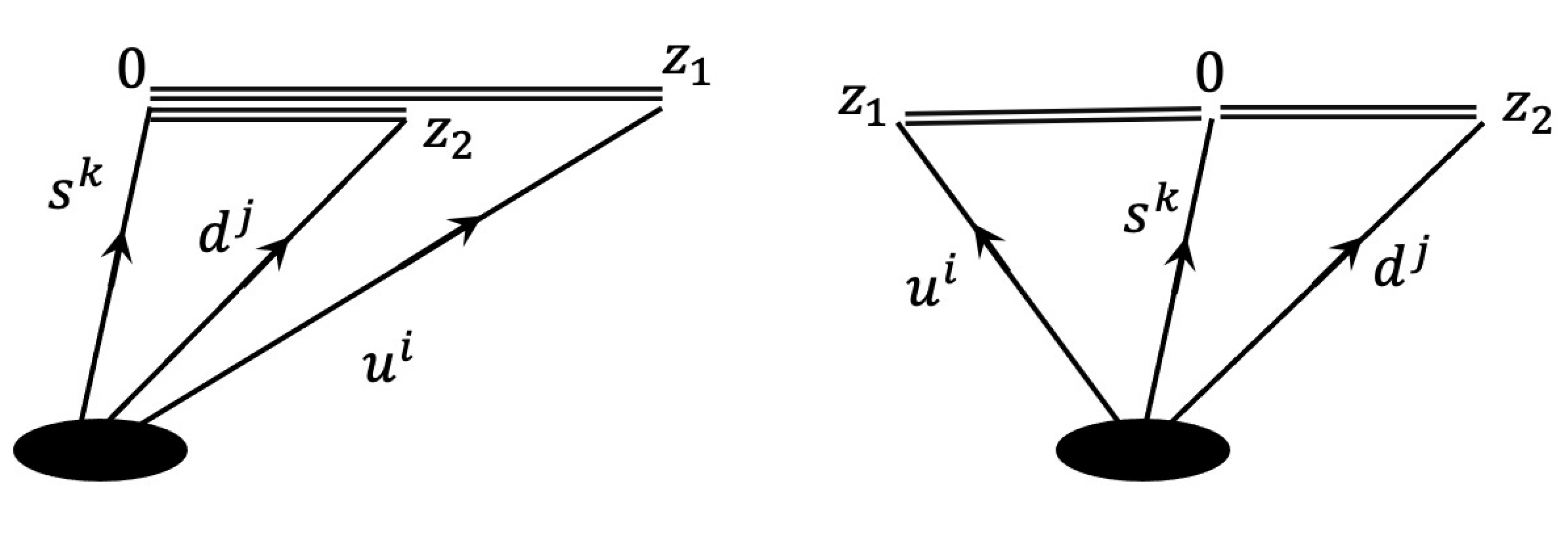}
\caption{The Wilson lines for two cases of $z_1$ and $z_2$. At the left they point in the same direction while at the right they point in opposite directions.}
\label{fig:wilson_dir}
\end{figure}

The numerical uncertainties for quasi-light-front correlations increases rapidly with growing separation, and a brute-force truncation of the Fourier transformation can lead to unphysical oscillations in momentum space. Hence, in conjunction with the hybrid scheme, a physically-grounded extrapolation for long distances has been recommended  \cite{Ji:2020brr}.
The physically-grounded model, based on the asymptotic behavior of the light baryon LCDAs in momentum space, is:
\begin{align}
\phi(x_1,x_2)=C_0x_1^{d_1}x_2^{d_1}(1-x_1-x_2)^{d_2},
\label{eq:asymtotic_mom}
\end{align}
where $C_0$ is a normalization factor, and $d_{1,2}$ are undetermined exponents.

Generally, for PDFs or meson LCDAs, we can perform an analytical Fourier transform of the asymptotic behavior and take the limit of large LF distance to obtain the extrapolation form in coordinate space. However, for baryon LCDAs, the analytical Fourier transform becomes complicated, and similar to perturbative forms in self renormalization, encounters divergency issues. In this work, we adopt a simplified approach: Take the numerical Fourier transformation from Eq. (\ref{eq:asymtotic_mom}), and then perform the fitting for quasi-DAs in coordinate space:
\begin{align}
&\tilde{\Phi}(z_1,z_2;d_1,d_2)=\int_0^1dx_1\int_0^1dx_2\;e^{ix_1z_1P^z}e^{ix_2z_2P^z}\nonumber\\
&\qquad\qquad\qquad\qquad\times C_0x_1^{d_1}x_2^{d_1}(1-x_1-x_2)^{d_2}.
\label{eq:extrapolation}
\end{align}
This form is applied only when $z_1$ and $z_2$ are large, and hence we utilize it only in the coordinate regions marked by the red and blue boxes in Fig.~\ref{fig:extra_area}.

The asymmetry between the red and blue blocks in the first and fourth quadrants of Fig.~\ref{fig:extra_area} is evident. This lack of symmetry is a result of the non-symmetrical distribution of good signals in the $z_1-z_2$  plane, which has implications for our extrapolation process. Based on the definition of the baryon quasi-distribution amplitude and the prior analytic discussions, we consider two Wilson lines, $W^{ii'}(\vec{x},\vec{x}+z_1n_z)$   and $W^{jj'}(\vec{x},\vec{x}+z_2n_z)$, which link the quark fields in coordinate space. These two Wilson lines can be oriented either in the same direction or in opposite directions.

\vskip 1 cm
\vskip 1 cm

\begin{itemize}
\item  As illustrated in Fig.~\ref{fig:sym_region}, the Wilson lines point in opposite directions in the second and fourth quadrants, as depicted in the right panel of Fig.~\ref{fig:wilson_dir}. Consequently, in such scenarios, the effective Wilson length can be simply obtained as $|z_1|+|z_2|$.
\item   In the first and third quadrants, as shown in the left panel of Fig.~\ref{fig:wilson_dir}, when $z_1$ and $z_2$ point in the same direction, one can demonstrate:
\begin{align}
    &\;\;\;\;\;W\left(z_1, 0\right)^{i^{\prime} i} W\left(z_2, 0\right)^{j^{\prime} j} \epsilon^{i j k} \nonumber \\
    &=W\left(z_1, z_2\right)^{i^{\prime} i^{\prime \prime}} W\left(z_2, 0\right)^{i^{\prime \prime} i} W\left(z_2, 0\right)^{j^{\prime} j} \epsilon^{i j k} \nonumber \\
    &=W\left(z_1, z_2\right)^{i^{\prime} i^{\prime \prime}} W^{\dagger}\left(z_2, 0\right)^{k k^{\prime}} \epsilon^{i^{\prime \prime} j^{\prime} k^{\prime}}.
\end{align}
Therefore, in this scenario, the effective length of the Wilson lines is given by $\rm{Max}(|z_1| ,|z_2|)$. 
 \end{itemize}

Hence, the effective length of the Wilson lines is greater in the second and fourth quadrants than in the first and third quadrants. This implies that the uncertainty is more significant in the second and fourth quadrants than in the first and third quadrants for the same $|z_1|$ and $|z_2|$. Additionally, the size of the quasi-distribution amplitude will decay more rapidly in the second and fourth quadrants. Consequently, the signal will decrease in a rhombic pattern, as illustrated by the orange line in Fig.~\ref{fig:extra_area}.

Considering the symmetries described in Eq.~(\ref{eq:sym_prop}), it is sufficient to extrapolate the triangular regions 1 and 2, as illustrated in Fig.~\ref{fig:sym_region}. By combining these symmetries with the Wilson line properties discussed in the previous paragraph, we can now elaborate on the extrapolation regions outlined in Fig.~\ref{fig:extra_area}.
\begin{itemize}
\item On the axis $z_1=-z_2$, represented by the green blocks in Fig.~\ref{fig:extra_area}, the imaginary components of the quasi-distribution amplitude are constrained to be zero. Therefore, only the real part of $\Phi(z_1,z_2)$ within the green blocks contributes to the extrapolation process.
\item The red blocks mark  the regions where $z_1\gg 0$ and $z_2\gg0$. In these regions, both $z_1$
 and $z_2$  need to be extrapolated. However, when the lengths are equal, i.e., $|z_1|=|z_2|$, the effective Wilson line length is significantly larger in region 2 compared to region 1. Consequently, the red blocks in region 2 initiate from a smaller $|z_{1,2}|$.
In contrast, the blue blocks denote scenarios with a finite $z_2$  while $z_1\gg0$, and only $
 z_1$ requires extrapolation in this context. Similarly, due to the varying effective Wilson line lengths in regions 1 and 2, the shapes of the blue blocks will differ between the two regions.
\end{itemize}

\begin{widetext}
\begin{center}
\begin{figure}
\centering
\includegraphics[scale=0.35]{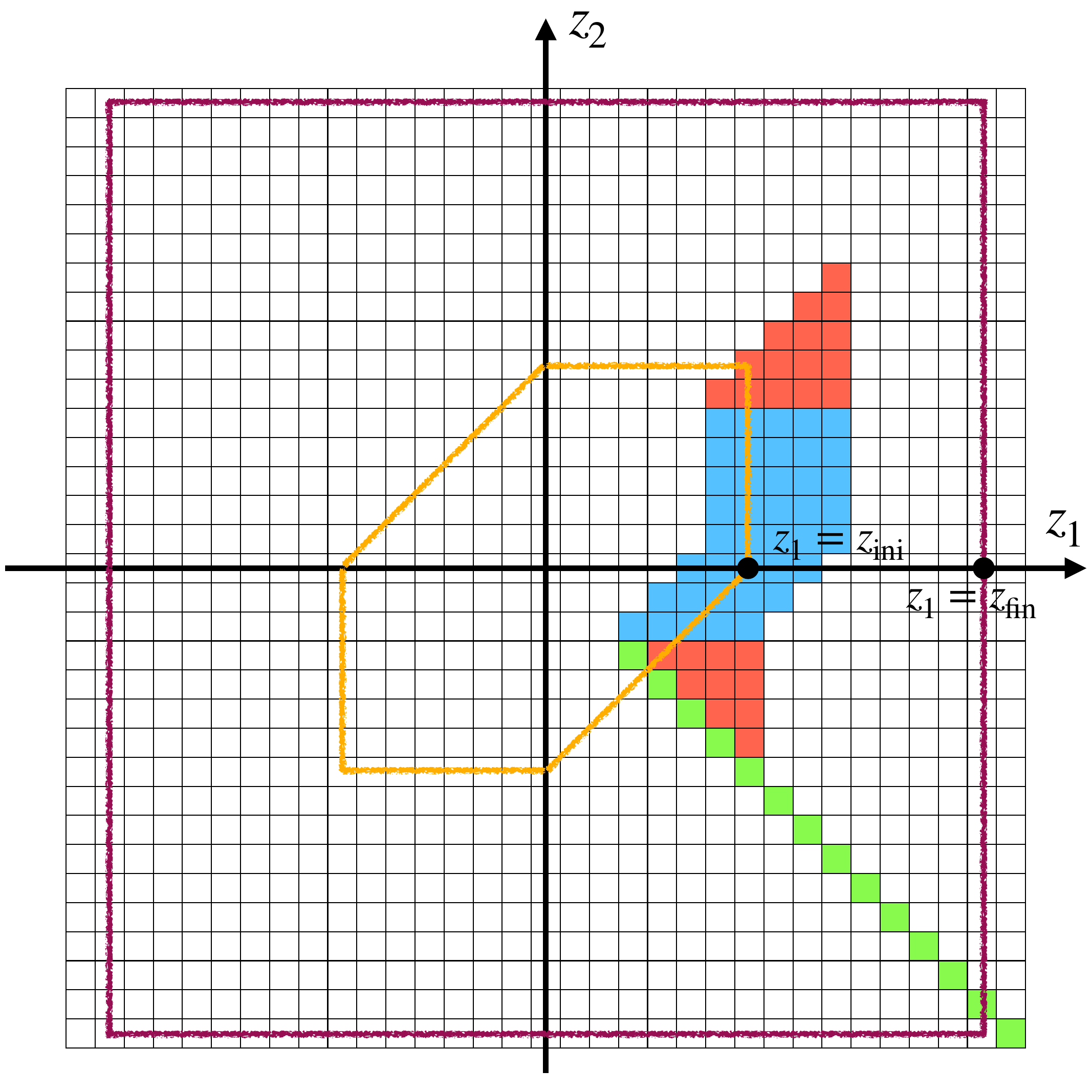}
\caption{The extrapolation regions in the $z_1-z_2$ plane. Raw data with good signal/noise were obtained for the rhombus indicated by the orange line. Extrapolations are based on these data. The meaning of the different color blocks is discussed in the main text. $z_{\rm ini}$ and $z_{\rm fin}$ indicate the region for which the quasi LCDA was extrapolated.}
\label{fig:extra_area}
\end{figure}
\end{center}
\end{widetext}



\subsection{Two dimensional matching}
\label{sec:2D_matching}

In LaMET, the quasi-distribution can be factorized into a perturbative matching kernel convoluted with a light-cone distribution. For the baryon LCDA involving two momentum fractions, this factorization relation extends to a two-dimensional scenario:
\begin{align}
\tilde{\phi}&(x_1,x_2)=\int_0^1dy_1\int_0^{1-y_1}dy_2C(x_1,x_2,y_1,y_2)\phi(y_1,y_2) \nonumber \\
&+\mathcal{O}\Big(\frac{\Lambda^2_{QCD}}{(x_1P^z)^2},\frac{\Lambda^2_{QCD}}{(x_2P^z)^2},\frac{\Lambda^2_{QCD}}{[(1-x_1-x_2)P^z]^2}\Big).
\label{eq:matching}
\end{align}
Here $\tilde{\phi}(x_1,x_2)$    represents the quasi-distribution amplitude in momentum space, and $\phi(x_1,x_2)$
 represents the LCDAs. The power correction terms   $\Lambda^2_{QCD}/(x_{1}P^z)^2$,  $\Lambda^2_{QCD}/(x_{2}P^z)^2$ and $\Lambda^2_{QCD}/[(1-x_1-x_2)P^z]^2$
 are suppressed in the large momentum limit. The matching kernel can be expressed through a perturbative expansion as:
\begin{align}
C(x_1,x_2&,y_1,y_2)=\delta(x_1-y_1)\delta(x_2-y_2)\nonumber\\
&+\frac{\alpha_sC_F}{2\pi}c^{(1)}(x_1,x_2,y_1,y_2)+\mathcal{O}(\alpha_s^2).
\label{eq:kernel}
\end{align}
The detailed expressions for $c^{(1)}(x_1,x_2,y_1,y_2)$ in terms of the "double-plus function" $f_{\oplus}$ are compiled in Appendix \ref{sec:matching_details}.

The  matching of the baryon distribution amplitude involves a two-dimensional matching process, as both the quasi-distribution amplitude and the light-cone distribution amplitude have two momentum fraction variables. Hence, the matching kernel comprises four variables. Two numerical methods are commonly used for implementing the matching:
\begin{itemize}
\item  Numerically discretizing the matching kernel and performing its inversion to achieve the matching.
\item Expanding the higher-order terms in the matching as perturbations  and iteratively solving for the matching.
\end{itemize}
In the matching process, the matching kernel is represented as a four-dimensional tensor. While it is possible to reduce this tensor to two dimensions by utilizing the properties of the double-plus function and subsequently carry out a numerical inversion, we introduce and employ an iterative approach in this work to facilitate a more straightforward numerical implementation for the matching procedure.

The dominant term in the matching kernel, as given by Eq.~(\ref{eq:kernel}), is a Dirac delta function. The discrepancy between $\phi$ and $\tilde{\phi}$  is of order $\alpha_s$ at the one-loop level. Thus, by substituting Eq.~(\ref{eq:kernel}) into Eq.~(\ref{eq:matching}), we obtain:
\begin{align}
\phi(x_1,x_2)&=\tilde{\phi}(x_1,x_2)-\frac{\alpha_sC_F}{2\pi}\int_{-\infty}^{\infty}dy_1\int_{-\infty}^{\infty}dy_2\nonumber\\
&\times c^{(1)}(x_1,x_2,y_1,y_2)\tilde{\phi}(y_1,y_2)+\mathcal{O}(\alpha_s^2).
\label{eq:matching_order}
\end{align}

In this equation, we extend the integration limits of $y_{1,2}$ from $0$ to $1$ and $0$ to $1-y_1$ to $-\infty$ to $\infty$, a modification that may introduce systematic uncertainties at higher orders. The subsequent leading-order matching kernel $c^{(1)}(x_1,x_2,y_1,y_2)$ is a double-plus function involving the variables $x_1$  and $x_2$:
\begin{align}
c^{(1)}&(x_1,x_2,y_1,y_2)=\left[f(x_1,x_2,y_1,y_2)\right]_{\oplus}\nonumber\\
&=f(x_1,x_2,y_1,y_2)-\delta(x_1-y_1)\delta(x_2-y_2)\nonumber\\
&\times\int_{-\infty}^{\infty}dt_1\int_{-\infty}^{\infty}dt_2f(t_1,t_2,y_1,y_2).
\label{eq:kernel_loop}
\end{align}
By substituting Eq.~(\ref{eq:kernel_loop}) into Eq.~(\ref{eq:matching_order}), these singularities in $f(x_1,x_2,y_1,y_2)$ can be canceled between the first and second term in Eq.~(\ref{eq:kernel_loop}) with the constraints $x_1=y_1$ and $x_2=y_2$. Then the matching equation simplifies to:
\begin{align}
\phi(x_1,&x_2)=\tilde{\phi}(x_1,x_2)-\frac{\alpha_sC_F}{2\pi}\Bigg[\int dy_1dy_2\Big[f(x_1,x_2,y_1,y_2)\nonumber\\
&\times\tilde{\phi}(y_1,y_2)-f(y_1,y_2,x_1,x_2)\tilde{\phi}(x_1,x_2)\Big]\Bigg],
\label{eq:matching_x1x2}
\end{align}
without any plus function. 

Recent studies have explored matching kernels in different renormalization schemes for the  baryon LCDA. $\overline{\rm MS}$ bar scheme and RI/MOM scheme matching kernels for the leading twist function $A(z_iP\cdot n)$ are presented in \cite{Deng:2023csv}. The hybrid scheme matching kernels for $V(z_iP\cdot n)$, $A(z_iP\cdot n)$ and $T(z_iP\cdot n)$ functions combining self-renormalization and ratio scheme renormalization have been collected in \cite{Han:2023xbl, Han:2024ucv}. Furthermore, the matching kernels in coordinate space in the ratio scheme for the three leading twist distritbutions are displayed in \cite{Han:2023hgy}. From all of them, the ratio scheme matching kernel is adopted in this work. For further details and discussions on the matching kernel, please refer to the Appendix. 


\section{Numerical results}
\subsection{Lattice setup}

The configurations used in this study are based on the stout smeared-clover fermion action coupled with Symanzik gauge action, under periodic boundary conditions, generated by the CLQCD collaboration \cite{CLQCD:2023sdb}. These configurations have been successfully applied in studies involving hadron spectrum~\cite{Liu:2022gxf,Xing:2022ijm,Yan:2024yuq}, decay and mixing of charmed hadron~\cite{Zhang:2021oja,Liu:2023feb,Liu:2023pwr,Meng:2024nyo,Du:2024wtr}, heavy meson LCDAs~\cite{Han:2024min,Han:2024yun}, PDFs  of a deuteron-like  dibaryon~\cite{Chen:2024rgi} and other interesting phenomena~\cite{Zhao:2022ooq,Meng:2023nxf}.  We focus on a single ensemble, F32P30, characterized by a lattice spacing of $a=0.07746(18)$ fm and a volume of $n_s^3 \times n_t = 32^3 \times 96$.  The valence light quark mass is set to match the sea quark mass, with $m_{\pi} = 303.2(1.3)$ MeV. To explore the momentum dependence of the $\Lambda$ LCDA and approach the infinite momentum limit, we consider three different hadron momenta: $P^z = \{5,6,7\}\times2\pi/(n_sa)=\{2.52, 3.02, 3.52\}$ GeV. Our numerical simulations rely on point sources with momentum smearing\cite{Bali:2016lva} and APE smearing\cite{APE:1987ehd} during make sources, techniques to significantly improve the signal-to-noise ratio in large $P_z$. 
The parameters in momentum smearing are taken as $p=3$ with smearing size and iteration number $\{5,50\}$. Overall, our simulation comprises 777 configurations, with $9 \times 6$ source positions on each configuration, leading to a total of $777 \times 9 \times 6$ measurements.


\subsection{Dispersion relation}

To analyze the dispersion relation, we evaluate the local two-point correlation function $C_2(0,0;t,P^z)$, as defined in Eq. (\ref{eq:2pt_definition}), across a range of hadron momenta from 0 to $3.52$ GeV. The effective energy is extracted by fitting the parameterization $C_2(0,0;\tau,P^z)=c_0e^{-E_0\tau}(1+c_1e^{-\Delta E\tau})$
for Euclidean time $\tau=it$. In this expression, the ground state energy $E_0$ characterizes the effective energy of the hadron at $P^z$.

\begin{figure}
\centering
\includegraphics[scale=0.65]{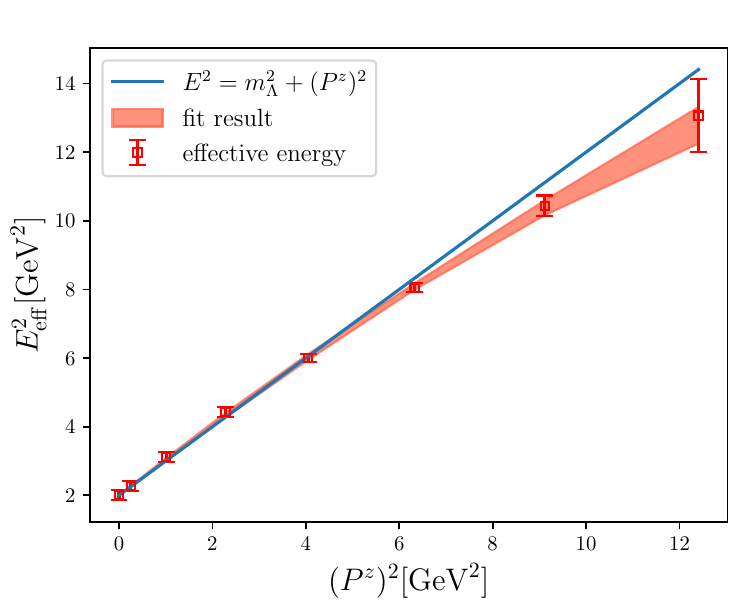}
\caption{Dispersion relation for the $\Lambda$ baryon on the ensemble F32P30.}
\label{fig:dispersion}
\end{figure}

To quantify the discretization effects in the relationship between the effective energy and hadron momentum, we utilize the following fitting ansatz:
\begin{align}
E^2 = m_{\Lambda}^2 + c_2 (P^z)^2 + c_3 (P^z)^4,
\end{align}
where the parameters $c_2$  and $c_3$ are chosen to quantify the deviation from the continuum relation, $E^2 = m_{\Lambda}^2 + (P^z)^2$. The fitted values obtained, as depicted in Fig.~\ref{fig:dispersion}, are $c_2 = 1.058(44)$ and $c_3 = -0.0152(66)$, which are in proximity of 1 and 0, respectively. Hence, a mild deviation is noted between the lattice outcome and the continuum prediction.

\subsection{Extraction of Quasi-DA}
\label{sec:2pt_qda_co}

\begin{figure}
\centering
\includegraphics[scale=0.65]{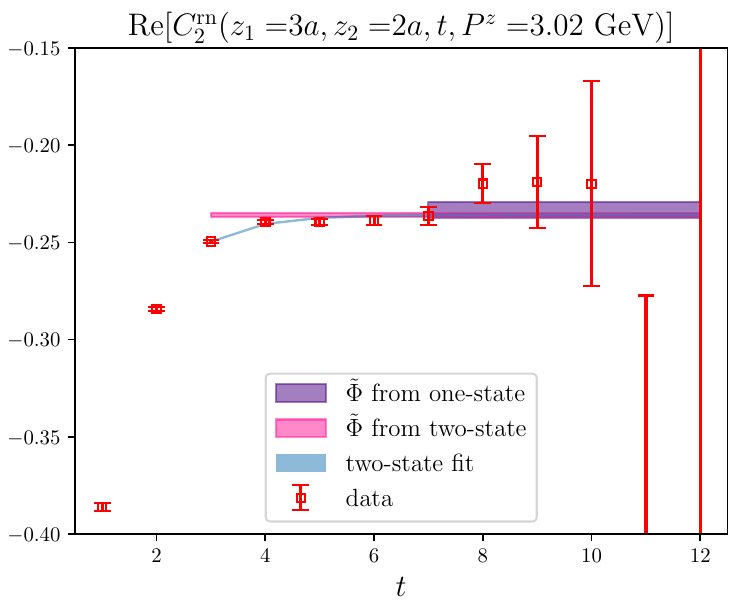}
\includegraphics[scale=0.65]{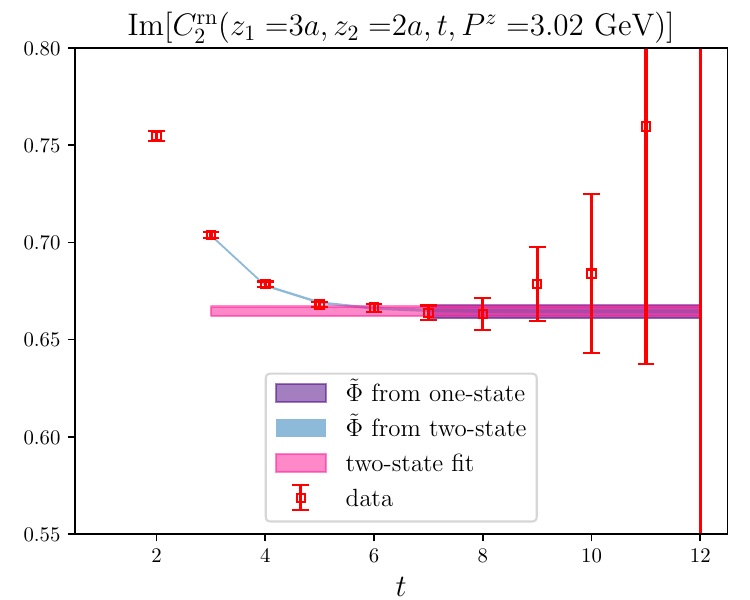}
\caption{Comparison of one- and two- state fit for the renormalized 2pt function. The upper and lower panels show the real and imaginary parts of $C_2^{\rm rn}(z_1,z_2,t,P^z)$ at $\{z_1,z_2,P^z\}=\{3a,2a,3.02\;\rm{GeV}\}$. In each figure the colored bands  for $\tilde{\Phi}$ represent the plateaus and the fit ranges.}
\label{fig:one_two_fit}
\end{figure}

In lattice simulations, the two-point correlation function $C_2(z_1,z_2;t,P^z)$ is utilized to determine the equal-time matrix element, as defined in Eq. (\ref{eq:matrix_element}), including the contribution from the ground state. To extract the ground state matrix element, which represents the quasi-distribution amplitude, we analyze the renormalized two-point function $C^{\rm rn}_2(z_1,z_2;t,P^z)$
\begin{align}
C^{\rm rn}_2(z_1,z_2&;t,P^z)=\frac{C_2(z_1,z_2;t,P^z)C_2(0,0;t,0)}{C_2(z_1,z_2;t,0)C_2(0,0;t,P^z)}\nonumber \\
&=\tilde{\Phi}(z_1,z_2,P^z)\frac{1+c_0(z,P^z)e^{-\Delta E\tau}}{1+c_1e^{-\Delta E\tau}}.
\end{align}
Here $\tau$ represents the Euclidean time, with $c_0$, $c_1$, and $\Delta E$ serving as parameters that account for excited state contamination. As $\tau$ becomes sufficiently large, the term $e^{-\Delta E \tau}$  significantly suppresses the contribution from excited states, thereby justifying a one-state fit at large $\tau$. To compare the effectiveness of one-state and two-state fits, we conduct an analysis at a specific instance of $z_1, z_2$, as shown in Fig.~\ref{fig:one_two_fit}. The results indicate that the one-state fit not only provides a clear interpretation of the data but also demonstrates greater robustness and stability across different values of $z_1$, $z_2$ , and $P^z$ when compared to the two-state fit. As a result, we will utilize the one-state fit for subsequent analyses.

\begin{figure}
\center
\includegraphics[scale=0.65]{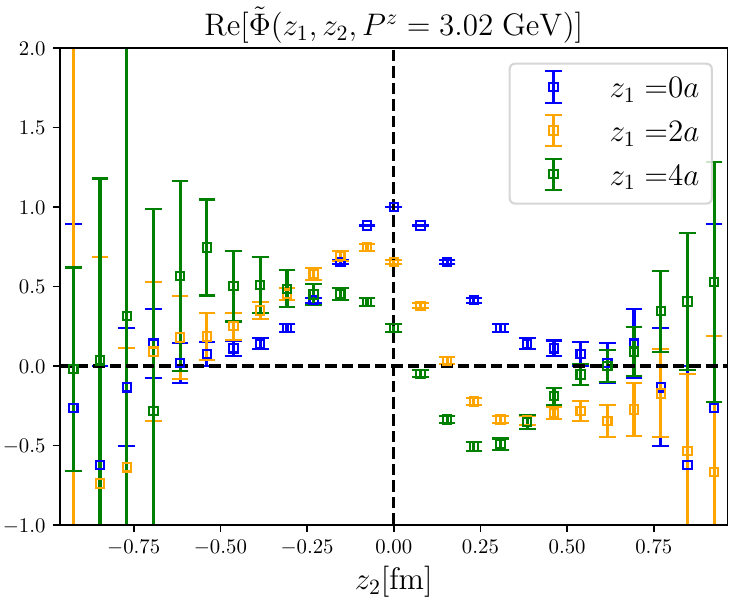}
\includegraphics[scale=0.65]{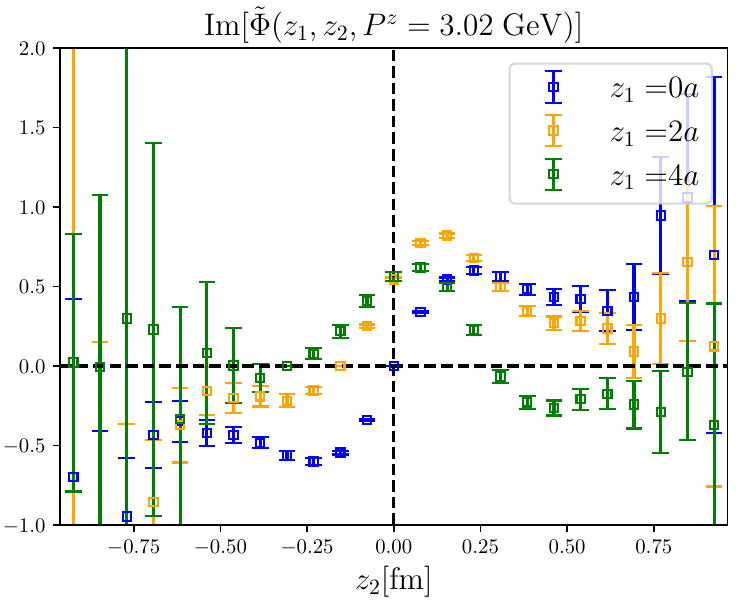}
\caption{The figures correspond to Re[$\tilde{\Phi}(z_1,z_2)$] (upper panel) and Im[$\tilde{\Phi}(z_1,z_2)$] (lower panel) with fixed $z_1$ at $P^z=3.02$ GeV.}
\label{fig:qda_co_pm}
\end{figure}

The renormalized quasi-distribution amplitude  of a baryon depends on both $z_1$ and $z_2$ . To elucidate the behavior of the quasi-DA as the non-local separation increases, we present $\tilde{\Phi}(z_1,z_2)$ in Fig. \ref{fig:qda_co_pm} and investigate the variation with $z_2$ for fixed $z_1$ at several specific values. When $z_1=0a$, the real and imaginary parts of $\tilde{\Phi}(z_1,z_2)$ exhibit axial and central symmetry at axis $z_2=0$ and point $z_1=z_2=0$ respectively. However, this symmetry diminishes as $z_1$ increases. Further examples of $\tilde{\Phi}(z_1,z_2)$ with fixed $z_1$ are discussed in Appendix~ \ref{sec:qda_co}, showing similar trends. Moreover, the data indicates slightly larger uncertainties in the region $z_2<0$
 compared to $z_2>0$, which can be attributed to the orientation of the Wilson lines for $z_{1,2}$ being either parallel or antiparallel, as depicted in Fig. \ref{fig:wilson_dir}. Additionally, the uncertainties notably increase at larger $z_2$ values due to the extended Wilson lines in non-local operators. This highlights the necessity of performing an extrapolation in both $z_1$ and $z_2$ to accurately capture the long-distance information, as elaborated in Sec. \ref{sec:extra}.


\begin{figure}
\center
\includegraphics[scale=0.7]{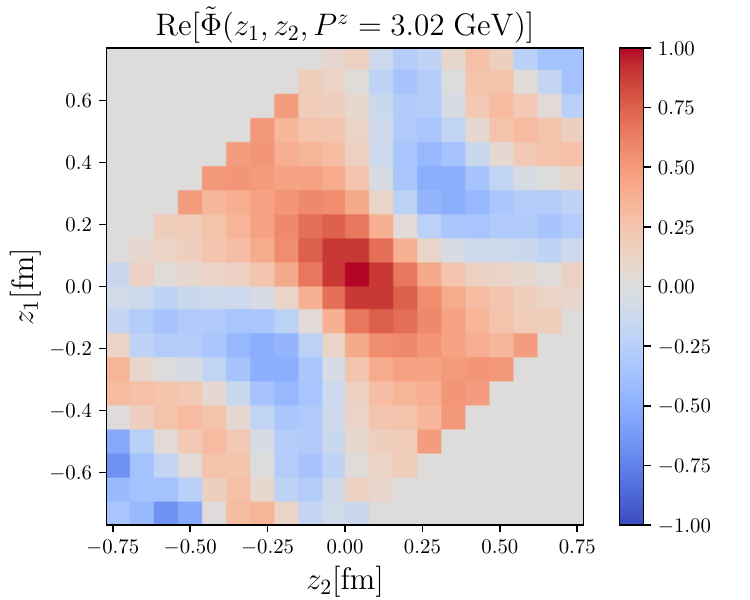}
\includegraphics[scale=0.7]{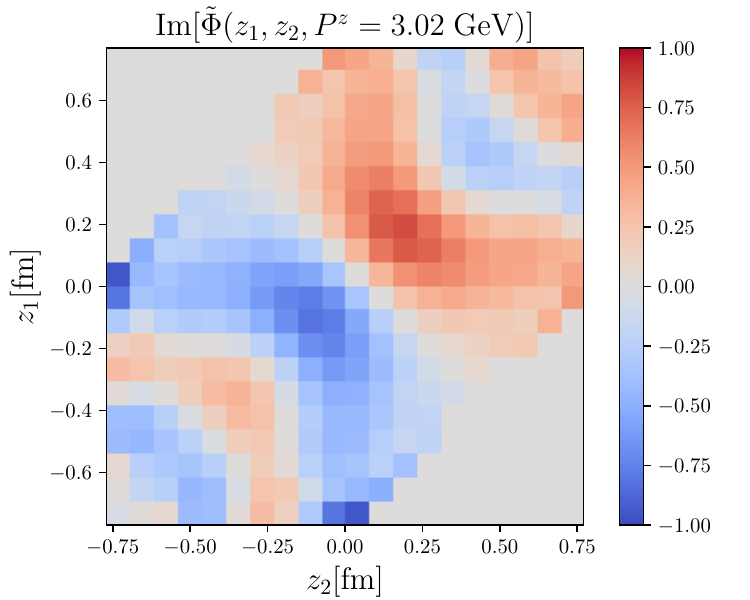}
\caption{The central value of $\tilde{\Phi}(z_1,z_2)$ in the $z_1$-$z_2$ plane is displayed in the figures, including the real (upper panel) and imaginary (lower panel) part. The color represents the value.}
\label{fig:qda_co_3D}
\end{figure}

To demonstrate the relationship between $z_1$ and $z_2$, Fig. \ref{fig:qda_co_3D} presents a three-dimensional visualizations of the central values of the renormalized quasi-distribution amplitude in coordinate space, $\tilde{\Phi}(z_1,z_2)$. 
However, due to substantial uncertainties that become prominent at larger values of $z_1$
  and $z_2$, only data in the central region is displayed.
  
\begin{figure}
\center
\includegraphics[scale=0.7]{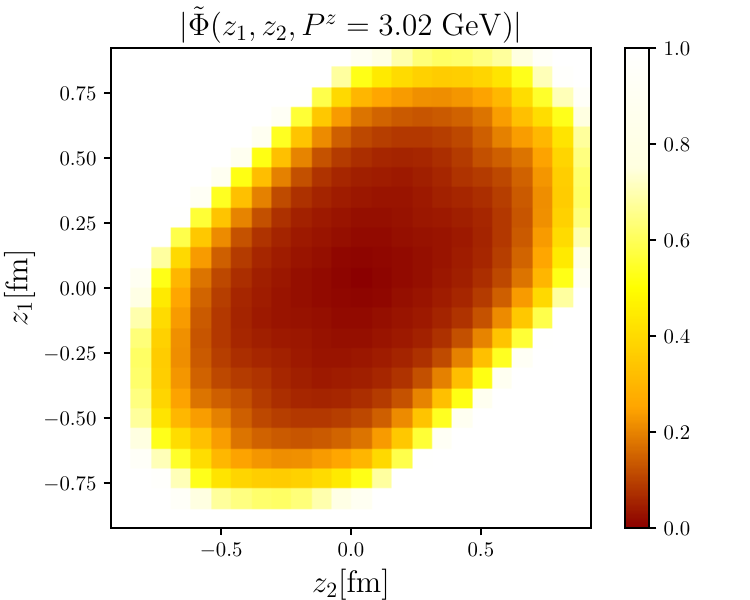}
\caption{The uncertainties of $|\tilde{\Phi}(z_1,z_2)|$ in the $z_1$-$z_2$ plane. A deeper color indicates a smaller uncertainty}
\label{fig:qda_co_3D_r}
\end{figure}

In this figure, several data characteristics align with the analytical expectations outlined in our framework:
\begin{itemize}
    \item Both the real and imaginary parts display pronounced oscillations along $z_1 = z_2$, with amplitudes that decay at a slow rate. This observation validates the expectation derived from the perturbative calculation discussed near Eq.~(\ref{eq:phi_co_per}).
    \item The imaginary part consistently approaches zero for $z_1 \rightarrow -z_2$, in accordance with the analytic properties dictated by the two symmetries.
    \item  A distinct signal is observed within a rhombic area, which correlates with the distribution of uncertainties depicted in Fig.~\ref{fig:qda_co_3D_r}. This pattern reflects the effective length for two Wilson lines with the same or opposite direction, as detailed in Sec.~\ref{sec:frame_extra}.
\end{itemize}

\subsection{Numerical Extrapolation}

Performing a Fourier transformation with  limited data points in coordinate space will  induce unphysical oscillations, as discussed in Sec. \ref{sec:extra}, and have been highlighted in previous studies \cite{Hua:2020gnw,LatticeParton:2022zqc,LatticePartonCollaborationLPC:2022myp,LatticeParton:2023xdl,LatticePartonLPC:2023pdv}. To eliminate these effects, we employ extrapolation techniques. According to Eq.~(\ref{eq:extrapolation}), we conduct a two-parameter fit for $d_1$ and $d_2$ applied to the renormalized matrix elements $\tilde{\Phi}(z_1, z_2, \mu, P^z)$, with the fitting region shown in Fig. \ref{fig:extra_area}. As explained in Sec. \ref{sec:extra}, the data points are grouped into red and blue blocks for fit procedure, corresponding to large $z_1\gg0$, $z_2\gg0$ and a finite $z_2$ but $z_1\gg0$, respectively.

Firstly, we examine the extrapolation results within the designated regions to assess the effectiveness of the fit function Eq.~(\ref{eq:extrapolation}) in both the blue and red regions as illustrated in Fig.~\ref{fig:extra_area}. The different results for the fitting parameters $d_1$ and $d_2$ are compiled in Tab.~\ref{tab:extra_paras_com}. These findings reveal that the values of $d_1$ and $d_2$ from analyses that include both regions, as well as those exclusively from the red region, are consistent within at most two standard deviations. This consistency demonstrates that the inclusion of the blue region exerts a minor influence. Consequently, we incorporate both the red and blue regions for extrapolation in the following analysis.

\begin{table}[htbp]
\centering
\caption{Fit parameters $d_1$ and $d_2$ obtained for the three hadron momenta $P^z=\{2.58,3.02,3.52\}$ GeV and utiliting in each case the data from either the red  or the red and blue region in Fig. \ref{fig:extra_area}.}
\label{tab:extra_paras_com}
\begin{tabularx}{0.49\textwidth}{>{\hsize=0.8\hsize\centering\arraybackslash}X|>{\hsize=1.1\hsize\centering\arraybackslash}X|>{\hsize=1.1\hsize\centering\arraybackslash}X}
\hline
\textbf{Momenta} & \textbf{$d_1$}, \textbf{$d_2$} from red region & \textbf{$d_1$}, \textbf{$d_2$} from red and blue regions \\ \hline
$P^z=2.52$ GeV & 0.042(40), 0.377(62) & 0.101(25), 0.459(39) \\ \hline
$P^z=3.02$ GeV & 0.049(45), 0.545(85) & 0.046(21), 0.480(45) \\ \hline
$P^z=3.52$ GeV & 0.072(55), 0.546(67) & 0.021(12), 0.449(30) \\ \hline
\end{tabularx}
\end{table}

Secondly, to mitigate the systematic effects due to our choice of fitting regions, we determine $d_1$ and $d_2$ for several slightly shifted fitting regions. The "standard" region encompasses the areas marked as red or blue in Fig.~\ref{fig:extra_area}. In addition, three other scenarios are analyzed in which this region is either shifted forward by one lattice unit in the $+z_1$ or $-z_1$ direction or by two units in the $+z_1$ direction. The maximal differences for $d_1$ and $d_2$ are treated as systematic error. Tab.~\ref{tab:extra_paras} summarizes the result including systematic error.

\begin{table}[htbp]
\centering
\caption{Final results for the fitting parameters $d_1$ and $d_2$.}
\label{tab:extra_paras}
\begin{tabularx}{0.49\textwidth}{>{\centering\arraybackslash}X|>{\centering\arraybackslash}X|>{\centering\arraybackslash}X}
\hline
\textbf{Momenta} & \textbf{$d_1$} & \textbf{$d_2$} \\ \hline
$P^z=2.52$ GeV & 0.118(68) & 0.50(22) \\ \hline
$P^z=3.02$ GeV & 0.052(34) & 0.49(18) \\ \hline
$P^z=3.52$ GeV & 0.13(12) & 0.63(21) \\ \hline
\end{tabularx}
\end{table}

\begin{figure}
\centering
\includegraphics[scale=0.65]{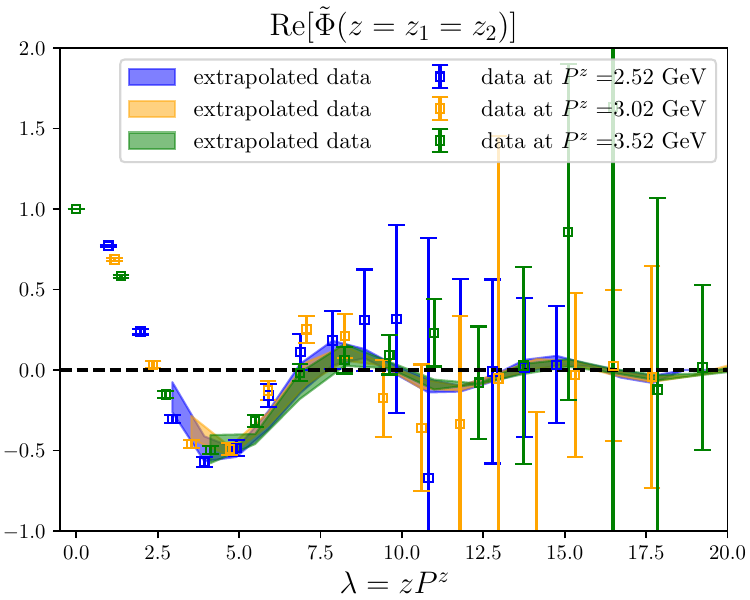}
\includegraphics[scale=0.65]{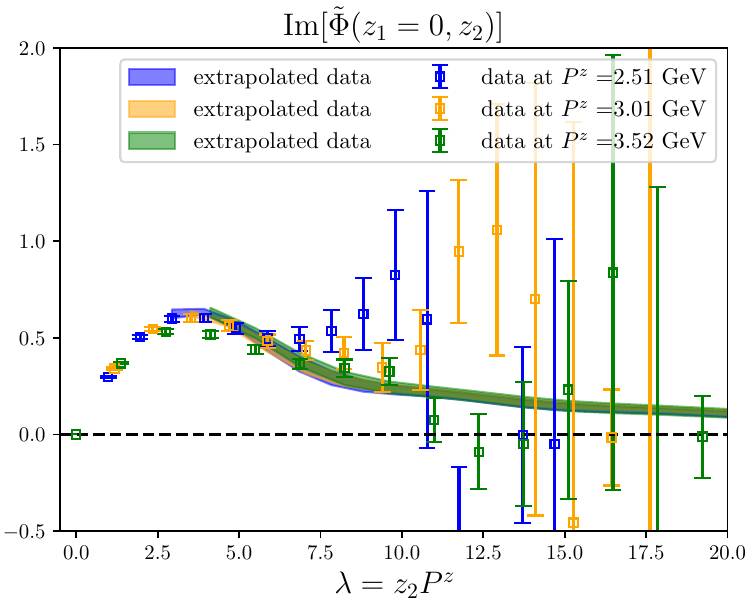}
\caption{The figures shows two examples of extrapolation for the renormalized matrix elements $\tilde{\Phi}(z_1,z_2,P^z)$ as functions of $z_1,z_2$. The upper panel corresponds to Re[$\tilde{\Psi}(\lambda=z_2P^z)$] when $z_1=z_2$ and the lower one represent Im[$\tilde{\Psi}(\lambda=z_2P^z)$] at $z_1=0$. The results of three hadron momenta are all displayed.  The fit region in each plots are approximately within $\lambda\sim[5,10]$, not all areas of colored bands.}
\label{fig:extra_co}
\end{figure}
Note that the central value of $d_1$ is smaller than that of $d_2$, and both are considerably smaller than unity. This discrepancy might lead to a mild convergence for $\tilde{\Phi}(z_1,z_2)$ as  $z_1$ and $z_2$ increase. Therefore, the endpoint behavior with $x_1,x_2\to 0$ in momentum space might also be affected. This situation could be improved in the future by performing a more precise calculation with larger statistics on a fine lattices. 

Finally, the extrapolated data are displayed in Fig. \ref{fig:extra_co}, where the upper panel illustrates the behavior of Re[$\tilde{\Phi}(z_1,z_2)$] in the red blocks shown in Fig. \ref{fig:extra_area}, and the lower panel corresponds to the blue blocks. As discussed in Sec.~\ref{sec:frame_extra}, the asymptotic behavior is applicable when $z_1$ and $z_2$ are large. In this analysis, the data included in the fitting predominantly falls within $\lambda \sim [5,10]$. Within this region, the extrapolated band aligns well with the original data, thereby validating the effectiveness of applying this asymptotic behavior, though more systematic uncertainties should be taken into account in future. 

As a result, we incorporate extrapolated data ranging from $z_{\rm ini}$ to $z_{\rm fin}$, as depicted in Fig.~\ref{fig:extra_area}. The regions within the gray lines are populated with lattice data, while the areas between the gray and blue lines are supplemented with data derived from extrapolation. This approach is justified as $\tilde{\Psi}(z_1, z_2)$ exhibits a clear signal within the inner hexagon region. 

\subsection{Quasi-DA in momentum space}
\begin{figure}
\centering
\includegraphics[scale=0.65]{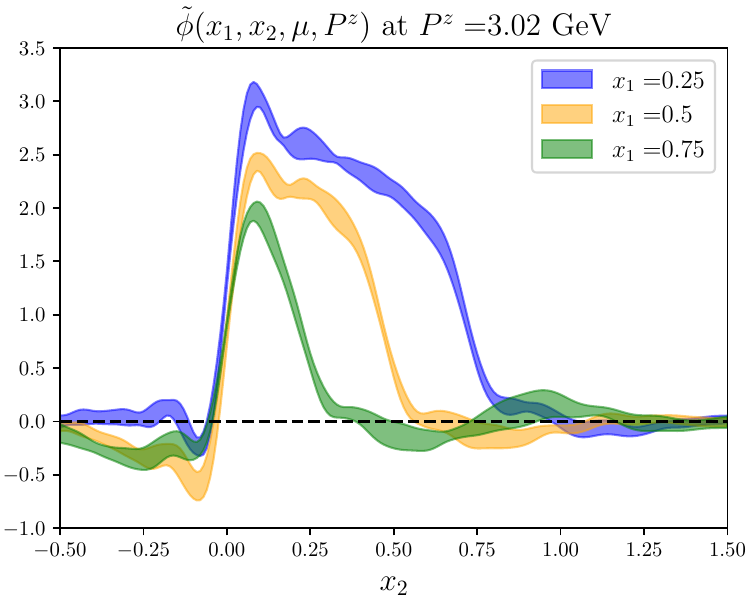}
\caption{The quasi-DA $\tilde{\phi}(x_1,x_2,\mu,P^z)$ in momentum space as a function of $x_2$ at $x_1=\{0.25,0.5,0.75\}$ with hadron momentum $P^z=3.02$ GeV. Because $\tilde{\phi}(x_1,x_2,\mu,P^z)$ is real by definition, only the real part is shown in this figure.}
\label{fig:qda_mom}
\end{figure}

The quasi-DA in momentum space $\tilde{\phi}(x_1,x_2,\mu,P^z)$, can be derived from the extrapolated quasi-DA in coordinate space $\tilde{\Phi}(z_1,z_2,\mu,P^z)$ by the Fourier transformation in Eq.~(\ref{eq:quasi_co}). We display the result in Fig.~\ref{fig:qda_mom}, where $\tilde{\phi}(x_1,x_2)$ is shown for $P^z = 3.02$ GeV and $x_1 = \{0.25, 0.5, 0.75\}$ as function of $x_2$. 

Since $x_1$ and $x_2$ are the momentum fractions of $u$ and $d$ quarks respectively, both of them are predominantly quite small, implying that the strange quark momentum fraction is quite large. This distribution can be partially  attributed to the SU(3) difference between the $u/d$ and $s$ quarks. Additionally, a small but non-vanishing tail of $\tilde{\phi}(x_1,x_2,\mu,P^z)$ appears in the unphysical region, defined as $\{(x_1,x_2) | x_1,x_2 < 0,\; \text{or}\; x_1,x_2 > 1,\; \text{or}\; x_1 + x_2 > 1\}$. Such behaviors might be caused by higher-twist contributions or other systematic uncertainties.

\subsection{Results for Baryon LCDAs and Phenomenological  Impact}

\begin{figure}
\centering
\includegraphics[scale=0.65]{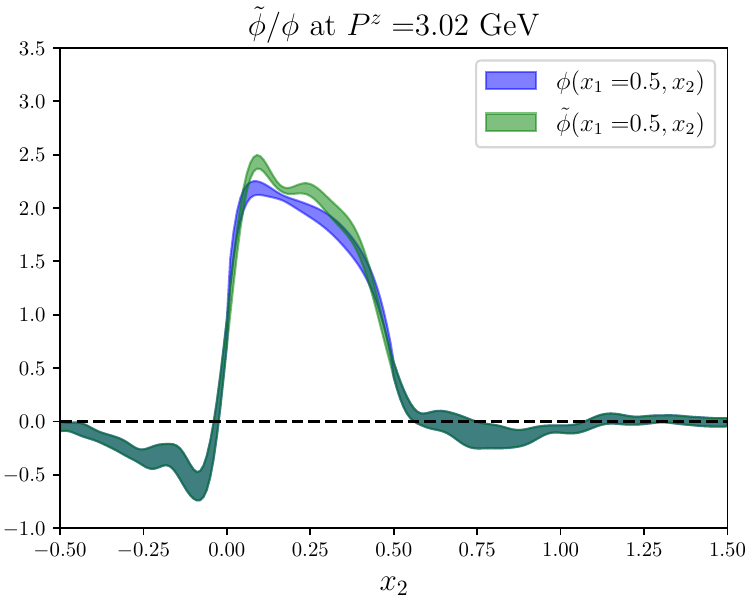}
\caption{Comparison of the $\Lambda$ LCDA $\phi(x_1,x_2,\mu,P^z)$ and the quasi-DA $\tilde{\phi}(x_1,x_2,\mu,P^z)$ at hadron momentum $P^z=3.02$ GeV as a function of $x_2$ at $x_1=0.5$.}
\label{fig:lcda_com}
\end{figure}

Direct implementation of the matching procedure for the two-dimensional baryon DA encounter great challenges. Specifically, one must address the double plus functions as defined in Eq.~(\ref{eq:kernel_loop}), which are likely to have divergences at several end points. Fortunately, by applying the properties of plus functions, one can simplify the matching formula in Eq.~(\ref{eq:matching_x1x2}). The formulation of the function $\left[f(x_1,x_2,y_1,y_2)\right]_{\oplus}$ in Eq.~(\ref{eq:kernel_loop}) incorporates three Dirac $\delta$-functions per term, thereby necessitating only a single integration for each term to determine $\phi(x_1, x_2)$ from $\tilde{\phi}(x_1, x_2)$. This streamlined approach is pursued in the subsequent analysis. More details are collected in Appendix.~\ref{sec:matching_details}.

Moreover, to remain the information of quasi-DA in the unphysical region, it is useful to apply the $\delta$-function in this region, together with one-loop corrections for the matching kernel in the physical region $\Lambda:\{(x_1,x_2) | 0 < x_1 < 1, \; 0 < x_2 < 1 - x_1\}$. Then the one-loop corrections becomes:
\begin{align}
\tilde{c}^{(1)}(x_1,x_2,y_1,y_2) = 
\begin{cases} 
c^{(1)}(x_1,x_2,y_1,y_2),{x_1,x_2\in\Lambda},\\
\delta(x_1-y_1)\delta(x_1-y_2),{x_1,x_2\notin\Lambda}.
\end{cases}
\end{align}

We then compare the quasi- and light-cone DAs to investigate the impact of one-loop corrections on the matching coefficient, as depicted in Fig.~\ref{fig:lcda_com}. The shown example has $x_1 = 0.5$ and $P^z = 3.02$ GeV, which corresponds to the orange band in Fig.~\ref{fig:qda_mom}. At leading order, the matching coefficient simplifies to a Dirac $\delta$-function, suggesting that the discrepancies observed between the quasi-DA and the light-cone DA are primarily attributable to one-loop corrections. Consequently, this analysis underscores that the one-loop corrections to the matching coefficient are not significant, but instrumental in attenuating the oscillations observed.

\begin{figure}
\centering
\includegraphics[scale=0.65]{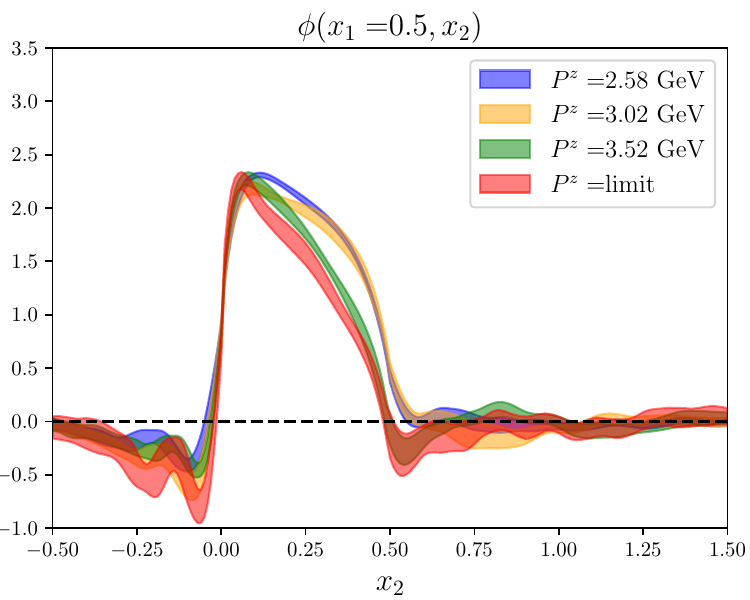}
\caption{Large momentum limit for the $\Lambda$ LCDA, including results at $P^z=\{2.58,3.02,3.58\}$ GeV and the extrapolated one.}
\label{fig:lcda_large_Pz}
\end{figure}

Next, we investigate the large momentum limit of our findings by analyzing the LCDA $\Phi(x_1, x_2, P^z)$ obtained at three distinct momenta. These values stem from the quasi-DA $\tilde{\Phi}(x_1, x_2, P^z)$ through the matching process outlined previously. To address the power corrections proportional to $1/(P^z)^2$, we perform an extrapolation as $P^z$ approaches infinity for $\phi(x_1,x_2,P^z)$ at three specific hadron momenta: $P^z = \{2.58, 3.02, 3.58\}$ GeV, as depicted by the equation:
\begin{align}
\phi(P_z) = \phi(P_z \to \infty) + \frac{c_2}{(P_z)^2} + \frac{c_4}{(P_z)^4}.
\end{align}
In this procedure we have quantified the systematic uncertainty as the variance between the LCDA at the highest momentum, $P^z = 3.58$
  GeV, and the limit as $P_z$  tends to infinity:
\begin{align}
\sigma_{\text{sys}} = |\phi(P^z \to \infty) - \phi(P^z = 3.58 \text{ GeV})|.
\label{eq:sys_pz}
\end{align}
The LCDA $\phi(x_1,x_2,P^z)$ outcomes at three hadron momenta, along with the extrapolated results, are depicted in Fig.~\ref{fig:lcda_large_Pz}. Notably, $\phi(x_1,x_2,P^z)$ at $P^z = 3.52$ GeV demonstrates a deviation from the results at lower momenta, $P^z = \{2.58, 3.02\}$ GeV.  

\begin{figure}
\centering
\includegraphics[scale=0.65]{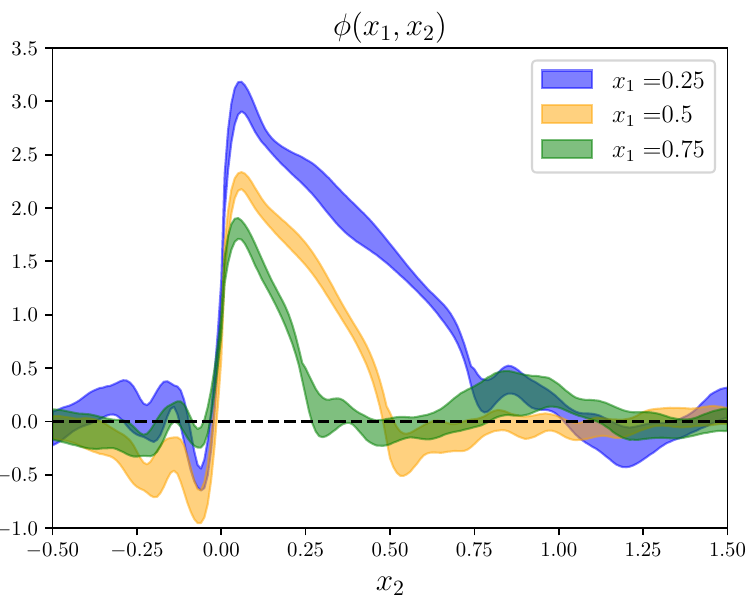}
\includegraphics[scale=0.7]{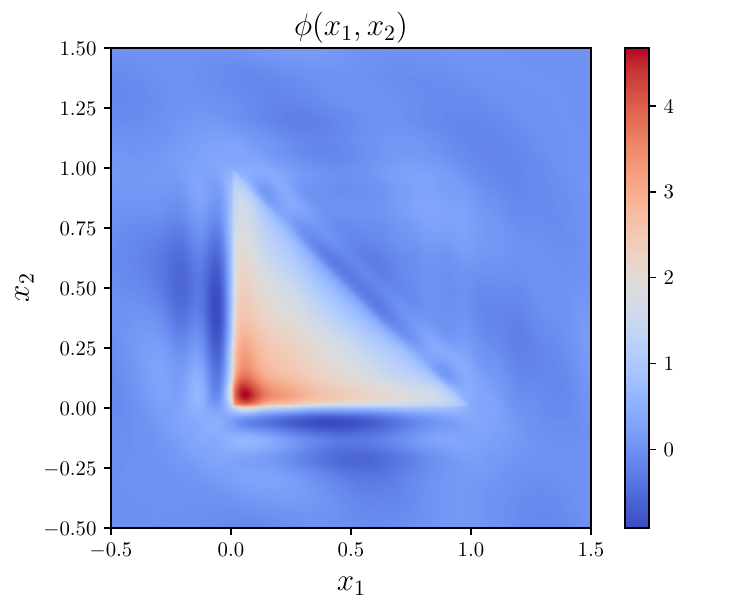}
\caption{The upper panel shows the $\Lambda$ LCDA at $x_1=0.25$, $0.5$, $0.75$ as a function of $x_2$ in the infinite momentum limit. The lower panel displays the two dimensional distributions of $x_1$ and $x_2$ but only central value of $\Lambda$ LCDA in infinite momentum limit. The unreliable region is estimated as $\{0<x_{1,2}<0.1,\;0.9<x_1<1,\;0.9-x_1<x_2<1-x_1\}$.}
\label{fig:lcda_final}
\end{figure}

The final results for the $\Lambda$ LCDA, $\phi(x_1,x_2)$, in the infinite momentum limit are showcased in Fig.~\ref{fig:lcda_final}. The upper panel of this figure displays $\phi(x_1,x_2)$ as a function of $x_2$ at fixed $x_1=0.25,0.5$, and $0.75$. The lower panel is a heat map to visually depict the distribution of LCDA across the $x_1-x_2$ plane. Statistical uncertainties are calculated from bootstrap samples of all measurements. Systematic uncertainties are attributed to various factors including the choice of fitting range for extrapolation, the selection of regions for utilizing extrapolated data, and the adoption of the infinite momentum limit.

The result of LCDA $\phi(x_1,x_2)$ refers to the leading twist LCDA $A$ for the $\Lambda$ baryon, which is illustrated in Fig.~\ref{fig:lcda_final}. Considering Eq.~(\ref{eq:matching}), owing to power corrections represented as $\Lambda^2_{QCD}/(x_{1,2}P^z)^2$ and $\Lambda^2_{QCD}/\left[(1-x_1-x_2)P^z\right]^2$, a region of unreliability emerges. This region is estimated to be $\{0<x_{1,2}<0.1,\;0.9<x_1<1,\;0.9-x_1<x_2<1-x_1\}$. According to Fig.~\ref{fig:lcda_final}, it exhibits two principal characteristics:
\begin{itemize}
    \item In the unphysical region, $\{x_2 < 0 \; \text{or} \; x_2 > 1 - x_1\}$, the LCDA $\phi(x_1,x_2)$ tends toward zero, likely influenced by the employment of large hadron momenta, the infinite momentum limit, and extensive $\lambda$ extrapolation.
    \item The LCDA $\phi(x_1,x_2)$ shows a dominant distributions in the regions where $x_1, x_2 \to 0$, aligning with observations in the quasi-DA. This might indicate a substantial contribution from the strange ($s$) quark to the momentum fraction in the $\Lambda$ baryon, or require a more systematic treatment of power corrections  at long distances, meriting further investigation.
\end{itemize}

As a phenomenological application, we apply our preliminary result to calculate the $\Lambda_b\to \Lambda$ form factors. We adopt the results using  lightcone sum rules \cite{Wang:2008sm} and calculate the form factor $g_2$.  To assist this calculation,  a parameterization of the $\Lambda$ LCDA is used by:
\begin{align}
\phi(x_1, x_2) = C x_1^{d_1} x_2^{d_1} (1 - x_1 - x_2)^{d_2},
\end{align}
where $d_1$, $d_2$ and $C$ are parameters to be fitted.  To optimize the results, we fit $\phi(x_1, x_2)$ within the medium region to avoid higher power corrections, yielding $d_1 = 0.90(36)$ and $d_2 = 1.59(40)$. 
It should be noted that this result obtained by fitting intermediate region of LCDA in momentum space differs significantly from the results in Tab.~\ref{tab:extra_paras} which by fitting coordinate space quasi-DA at large $z$. This also implies that the parameterization model is not entirely reasonable. For the LCDA results obtained from LaMET, it is preferable to directly substitute the numerical results.
Implementing these values in the calculation of the form factor, we obtain $g_2 = -0.0169(31)$ at Borel mass $M^2_B = 3\;\rm{GeV}^2$. Compared to the result in Ref.~\cite{Wang:2008sm} $g_2 = -0.02364$ where an asymptotic form is used for the LCDA  $\phi(x_1, x_2) = 120 x_1 x_2 (1 - x_1 - x_2)$, our findings on the leading twist LCDAs $\phi_A$  can make an reduction  of the form factor by approximately 20\%. This might have an impact on weak decays like $\Lambda_b \to \Lambda + l^+ + l^-$, where previous theoretical predictions overshoot the experimental data on $\Lambda_b\to \Lambda \mu^+\mu^-$ \cite{LHCb:2015tgy}.  These changes from the baryon LCDA might help resolve the tensions between experimental measurements and theoretical prediction, though this is very preliminary. 

\section{Conclusion and Prospects}

We have presented the first attempt to calculate the baryon light-cone distribution amplitude using LaMET on the lattice, taking the structure of a major contribution in the leading twist of $\Lambda$ baryon as an example. We discuss the challenges faced by the hybrid scheme in baryon distribution amplitudes and employed a simplified ratio as a preliminary attempt. Additionally, we simplify the parameterization used for the large $\lambda$ extrapolation. For the lattice numerical simulations, we utilize at one configuration with lattice spacing of 0.077 fm and pion mass of 303 MeV, performing calculations at three different momentum values for large momentum extrapolation.

From our numerical results, we observe that the quasi-DA in coordinate space aligns with our expectations regarding its analytic properties. The final distribution of the LCDA also reflects the fundamental characteristics of baryons. However, there are some issues  in the numerical results to be improved in future, such as renormalization scheme. Furthermore, our results represent only a non-physical pion mass with only one lattice spacing calculation, lacking comprehensive estimates of discretization errors and other systematic uncertainties.

Therefore, to achieve accurate and precise calculations of the baryon LCDA in future studies, we need to employ more configurations with smaller lattice spacings and higher statistics, while also adopting a more robust renormalization scheme and numerical analysis processes. A reliable computation of the baryon LCDA will significantly enhance phenomenological studies.

\section*{Acknowledgement} 
We thank Yu Jia and  Yu-Ming Wang for useful discussions.  We thank the CLQCD collaborations for providing us the gauge configurations with dynamical fermions~\cite{CLQCD:2023sdb}, which are generated on the HPC Cluster of ITP-CAS, the Southern Nuclear Science Computing Center(SNSC), the Siyuan-1 cluster supported by the Center for High Performance Computing at Shanghai Jiao Tong University and the Dongjiang Yuan Intelligent Computing Center. 
This work is supported in part by Guangdong Major Project of Basic and Applied Basic Research No. 2020B0301030008, Natural Science Foundation of China under grant No. 12125503, 12335003, 12375069, 12105247, 12275277, 12205106, 12175073, 12222503, 12147140, 12205180, 12375080, 1197505, 12293060, 12293062, and 12047503. J.H.Z is also supported by  CUHK-Shenzhen under grant No. UDF01002851. A.S., W.W., Y.Y., and J.H.Z. are also partially supported in part by a NSFC-DFG joint grant under grant No. 12061131006 and SCHA 458/22. J.Z.is also partially supported by the Project funded by China Postdoctoral Science Foundation under Grant No. 2022M712088. The computations in this paper were run on the Siyuan-1 cluster supported by the Center for High Performance Computing at Shanghai Jiao Tong University, and Advanced Computing East China Sub-center. The LQCD simulations were performed using the Chroma software suite~\cite{Edwards:2004sx} and QUDA~\cite{Clark:2009wm,Babich:2011np,Clark:2016rdz} through HIP programming model~\cite{Bi:2020wpt}. This work was partially supported by SJTU Kunpeng\&Ascend Center of Excellence.

\appendix
\begin{widetext}

\section{Detailed expressions for matching kernel}
\label{sec:matching_details}

The next-to-leading order of ratio scheme matching kernel $c^{(1)}(x_1,x_2,y_1,y_2)$ is given by the "double plus function" $f_{\oplus}$:
\begin{align}
c^{(1)}(x_1,x_2,y_1,y_2)&=\left[f(x_1,x_2,y_1,y_2)\right]_{\oplus}\nonumber\\
&=f(x_1,x_2,y_1,y_2)-\delta(x_1-y_1)\delta(x_2-y_2)\int_{-\infty}^{\infty}dt_1\int_{-\infty}^{\infty}dt_2f(t_1,t_2,y_1,y_2),
\label{eq:plus}
\end{align}
with
\begin{align}
f(x_1,x_2,&y_1,y_2)=\delta(x_2-y_2)\left[\frac{1}{4}Q_2(x_1,x_2,y_1,y_2)+\frac{7}{8}\frac{1}{|x_1-y_1|}\right]+\delta(x_1-y_1)\left[\frac{1}{4}Q_2(x_2,x_1,y_2,y_1)+\frac{7}{8}\frac{1}{|x_2-y_2|}\right]\nonumber\\
&\;+\delta(x_1+x_2-y_1-y_2)\left[\frac{1}{4}Q_3(x_1,x_2,y_1,y_2)+\frac{1}{4}Q_3(x_2,x_1,y_2,y_1)+\frac{3}{2}\frac{1}{|x_1-x_2-y_1+y_2|}\right],
\label{eq:c1}
\end{align}
where $f(x_1,x_2,y_1,y_2)$ is divided into three terms, each of which includes a Dirac $\delta$-function. As displayed in Fig.~\ref{fig:three_terms}, these three terms correspond to the one-loop contributions of two pairs of quarks exchanging gluons, where the spectator quark is represented by the Dirac $\delta$-function.

\begin{figure}
\centering
\includegraphics[scale=0.8]{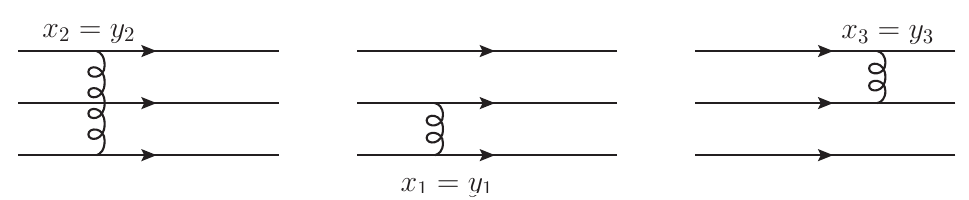}
\caption{The Feynman diagram correspond to three terms in one-loop matching coefficient.}
\label{fig:three_terms}
\end{figure}

The functions $Q_2$ and $Q_3$ are:
\begin{align}
& Q_2\left(x_1, x_2, y_1, y_2\right)= \nonumber\\
& \left\{\begin{array}{l}
\frac{\left(x_1+y_1\right)\left(x_3+y_3\right) \ln \frac{y_1-x_1}{-x_1}}{y_1y_3\left(y_1-x_1\right)}-\frac{x_3\left(x_1+y_1+2 y_3\right) \ln \frac{x_3}{-x_1}}{y_3\left(y_1-x_1\right)\left(y_1+y_3\right)}, \quad x_1<0 \\
\frac{\left(x_1-3 y_1-2 y_3\right) x_1}{y_1\left(x_3-y_3\right)\left(y_1+y_3\right)}-\frac{\left[\left(x_3-y_3\right)^2-2 x_3 y_1\right] \ln \frac{x_3-y_3}{x_3}}{\left(x_3-y_3\right) y_1y_3}+\frac{2 x_1 \ln \frac{4 x_1\left(x_3-y_3\right) p_z^2}{\mu^2}}{y_1\left(x_3-y_3\right)}+\frac{x_1 \ln \frac{4 x_1 x_3 p_z^2}{\mu^2}}{y_1\left(y_1+y_3\right)}, \quad 0<x_1<y_1 \\
\frac{\left(x_3-2 y_1-3 y_3\right) x_3}{y_3\left(x_1-y_1\right)\left(y_1+y_3\right)}-\frac{\left[\left(x_1-y_1\right)^2-2 x_1 y_3\right] \ln \frac{x_1-y_1}{x_1}}{\left(x_1-y_1\right) y_1 y_3}+\frac{2 x_3 \ln \frac{4 x_3\left(x_1-y_1\right) p_z^2}{\mu^2}}{y_3\left(x_1-y_1\right)}+\frac{x_3 \ln \frac{4 x_1 x_3 p_z^2}{\mu^2}}{y_3\left(y_1+y_3\right)}, \quad y_1<x_1<y_1+y_3 \\
\frac{\left(x_1+y_1\right)\left(x_3+y_3\right) \ln \frac{y_3-x_3}{-x_3}}{y_1 y_3\left(y_3-x_3\right)}-\frac{x_1\left(x_3+2 y_1+y_3\right) \ln \frac{x_1}{-x_3}}{y_1\left(y_3-x_3\right)\left(y_1+y_3\right)}, \quad x_1>y_1+y_3,
\end{array}\right.
\label{eq:Q2}
\end{align}

\begin{align}
& Q_3\left(x_1, x_2, y_1, y_2\right)= \nonumber\\
& \left\{\begin{array}{l}
\frac{\left(x_1 x_2+y_1 y_2\right) \ln \frac{x_2-y_2}{x_2}}{y_1y_2\left(x_2-y_2\right)}-\frac{x_1\left(x_2+y_1\right) \ln \frac{-x_1}{x_2}}{y_1\left(x_2-y_2\right)\left(y_1+y_2\right)}, \quad x_1<0 \\
\frac{1}{x_1-y_1}+\frac{2 x_1+x_2}{y_1\left(y_1+y_2\right)}+\frac{\left[\left(x_1+y_2\right) y_1-x_1^2\right] \ln \frac{x_2-y_2}{x_2}}{y_1y_2\left(x_2-y_2\right)}+\frac{x_1 \ln \frac{4 x_1\left(x_2-y_2\right) p_z^2}{\mu^2}}{y_1\left(x_2-y_2\right)}+\frac{x_1 \ln \frac{4 x_1 x_2 p_z^2}{\mu^2}}{y_1\left(y_1+y_2\right)}, \quad 0<x_1<y_1 \\
\frac{1}{x_2-y_2}+\frac{x_1+2 x_2}{y_2\left(y_1+y_2\right)}+\frac{\left[\left(x_2+y_1\right) y_2-x_2^2\right] \ln \frac{x_1-y_1}{x_1}}{y_1 y_2\left(x_1-y_1\right)}+\frac{x_2 \ln \frac{4 x_2\left(x_1-y_1\right) p_z^2}{\mu^2}}{y_2\left(x_1-y_1\right)}+\frac{x_2 \ln \frac{4 x_1 x_2 p_z^2}{\mu^2}}{y_2\left(y_1+y_2\right)}, \quad y_1<x_1<y_1+y_2 \\
\frac{\left(x_1 x_2+y_1 y_2\right) \ln \frac{x_1-y_1}{x_1}}{y_1y_2\left(x_1-y_1\right)}-\frac{x_2\left(x_1+y_2\right) \ln \frac{-x_2}{x_1}}{y_2\left(x_1-y_1\right)\left(y_1+y_2\right)}, \quad x_1>y_1+y_2,
\end{array}\right.
\label{eq:Q3}
\end{align}

\section{Quasi-DA in coordinate space}
\label{sec:qda_co}

\begin{figure}
\centering
\includegraphics[scale=0.63]{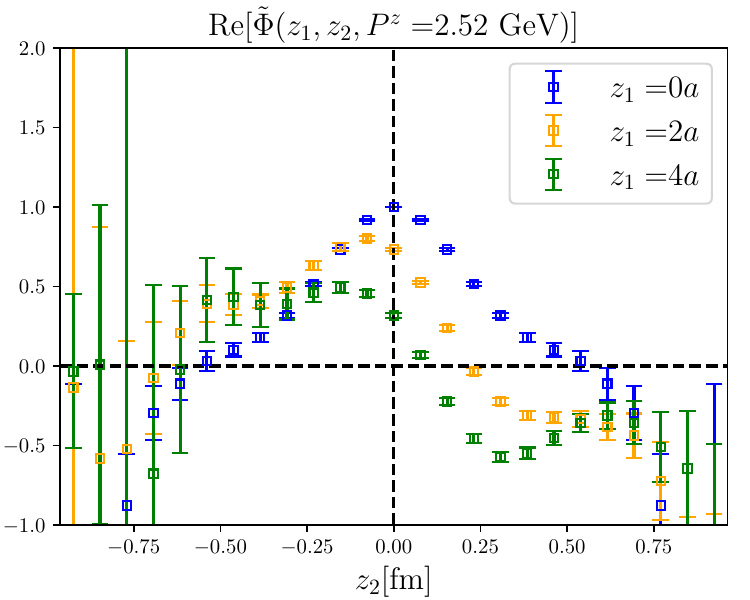}
\includegraphics[scale=0.63]{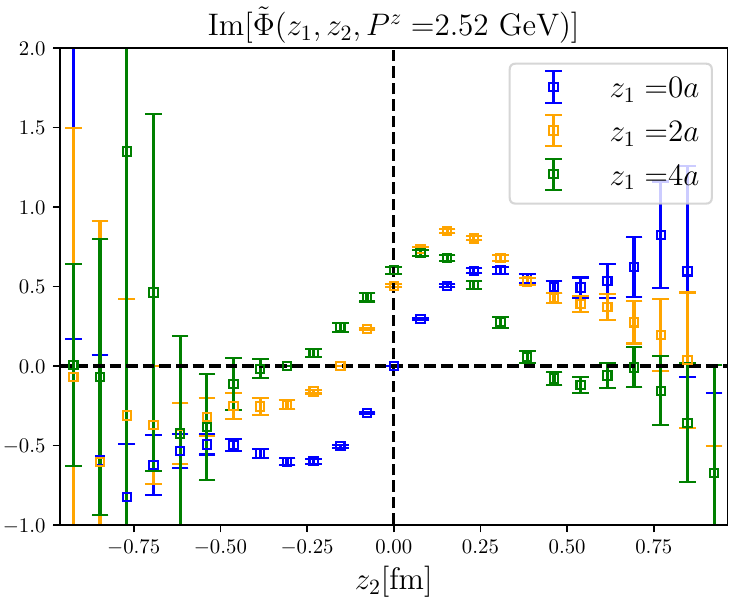}
\includegraphics[scale=0.63]{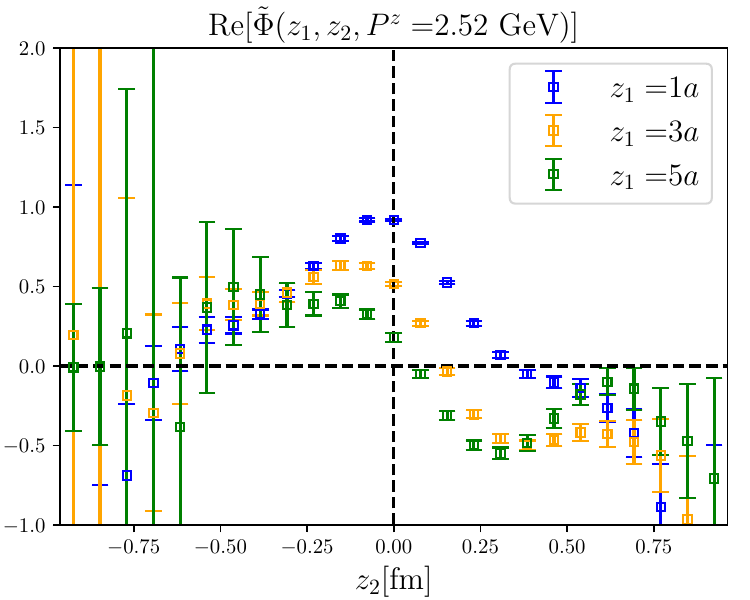}
\includegraphics[scale=0.63]{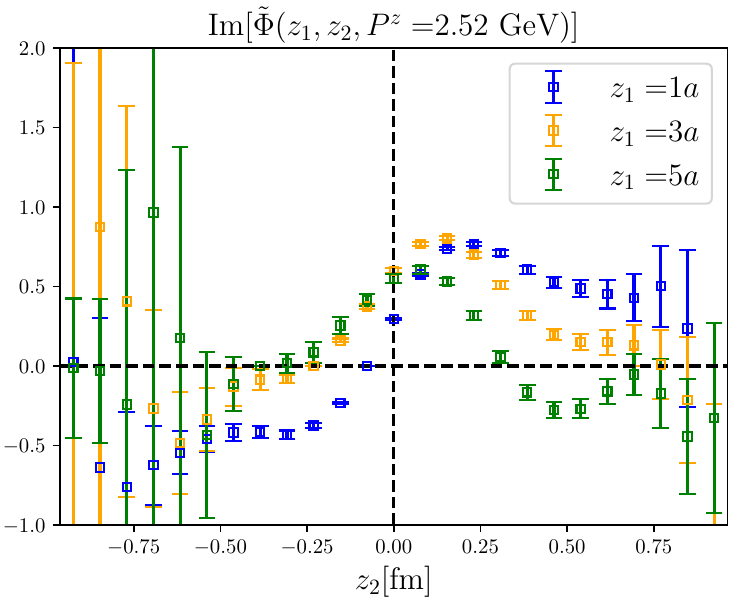}
\includegraphics[scale=0.63]{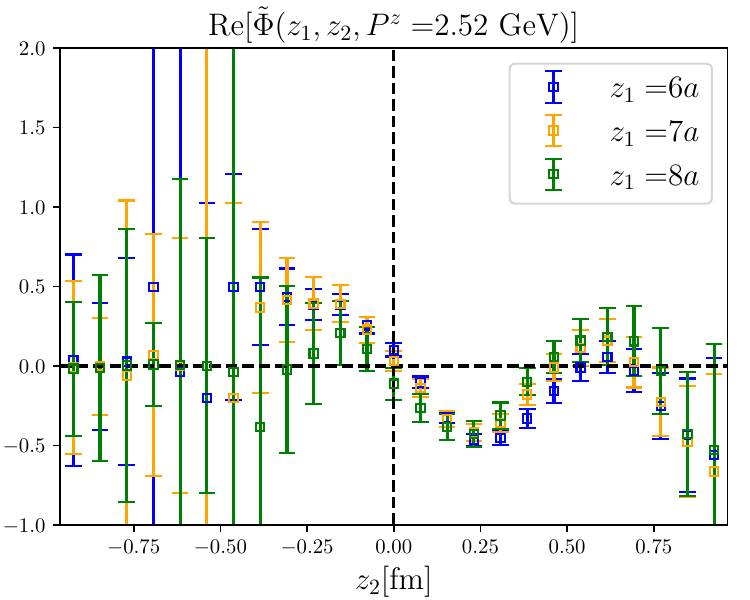}
\includegraphics[scale=0.63]{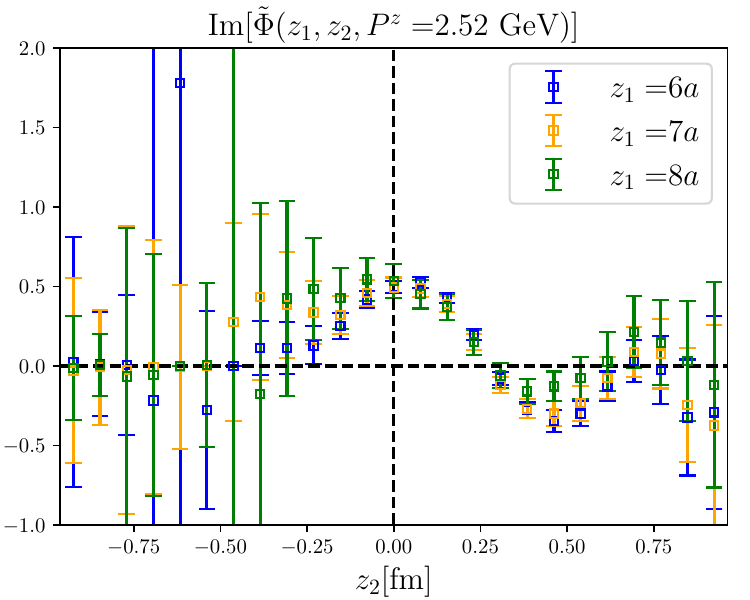}
\caption{The figures show the behavior of $\tilde{\Phi}(z_1,z_2)$ at all fixed $z_1\leq8a$ as a function of $z_2$. These are cases at $P^z=2.58$ GeV.}
\label{fig:qda_co_all_P5}
\end{figure}

\begin{figure}
\centering
\includegraphics[scale=0.63]{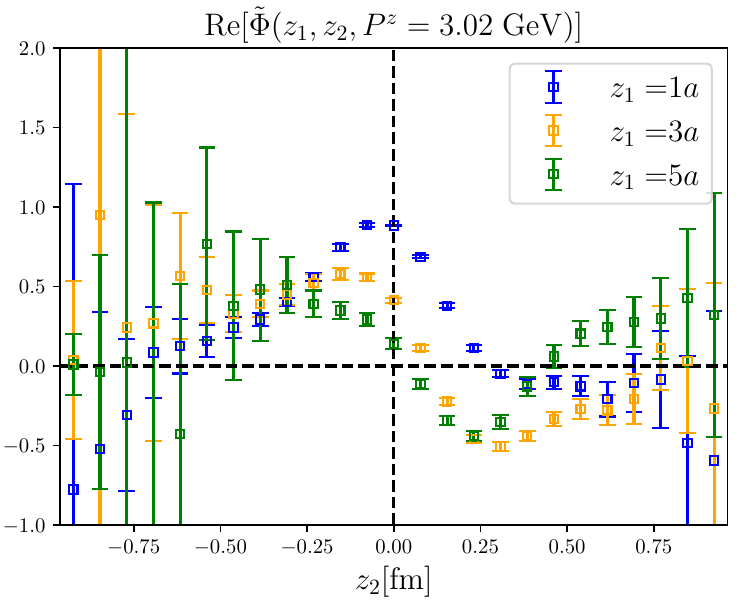}
\includegraphics[scale=0.63]{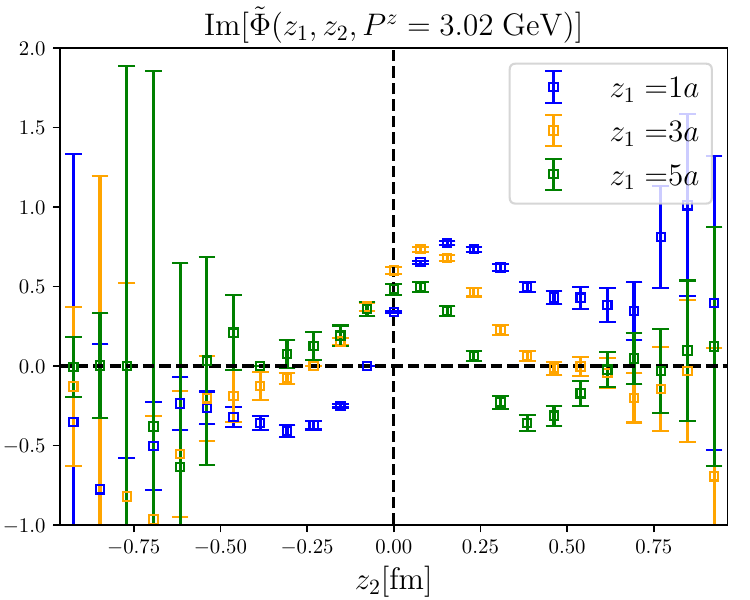}
\includegraphics[scale=0.63]{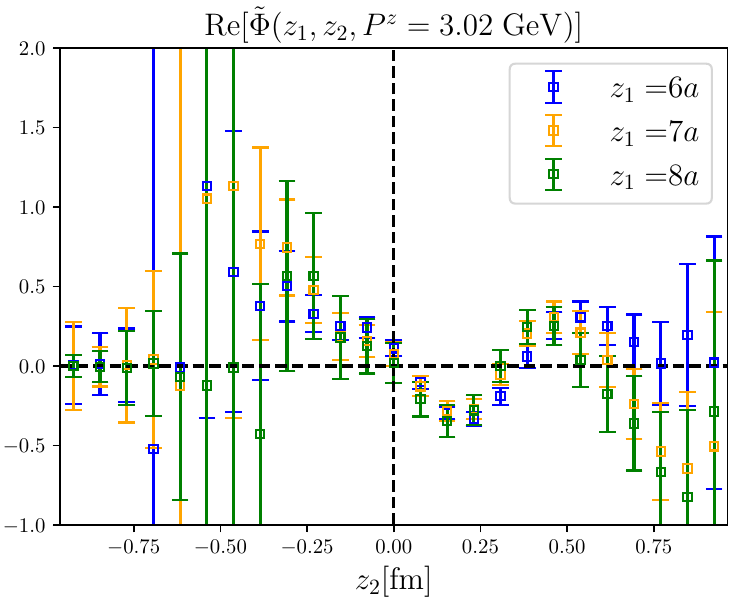}
\includegraphics[scale=0.63]{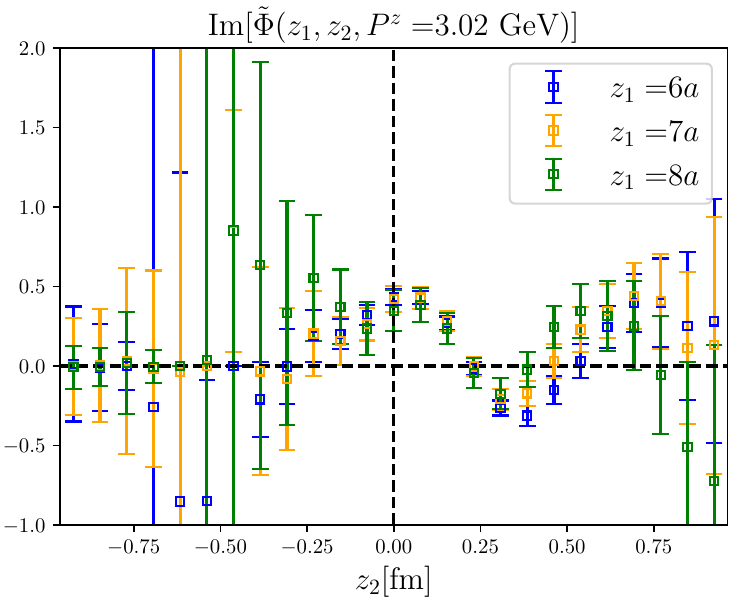}
\caption{The figures show the behavior of $\tilde{\Phi}(z_1,z_2)$ at all fixed $z_1\leq8a$ as a function of $z_2$. These are supplement cases to Fig. \ref{fig:qda_co_pm} in the main text at $P^z=3.02$ GeV.}
\label{fig:qda_co_all_P6}
\end{figure}

\begin{figure}
\centering
\includegraphics[scale=0.63]{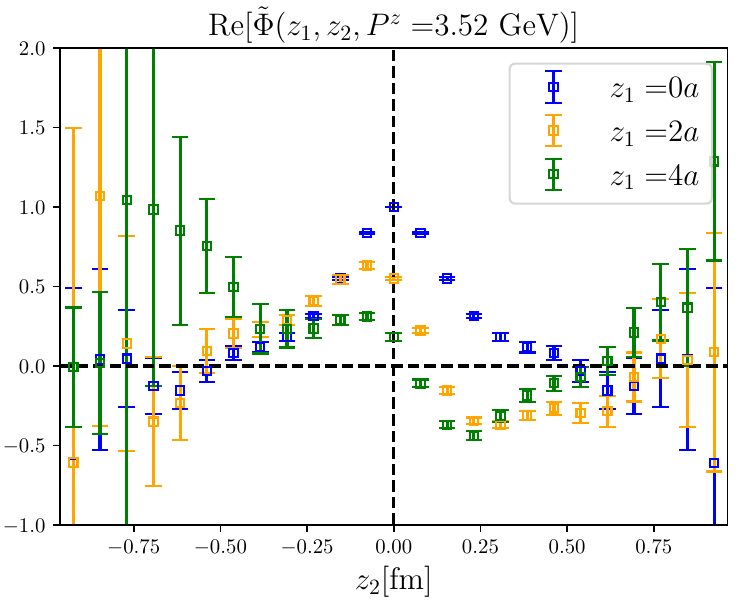}
\includegraphics[scale=0.63]{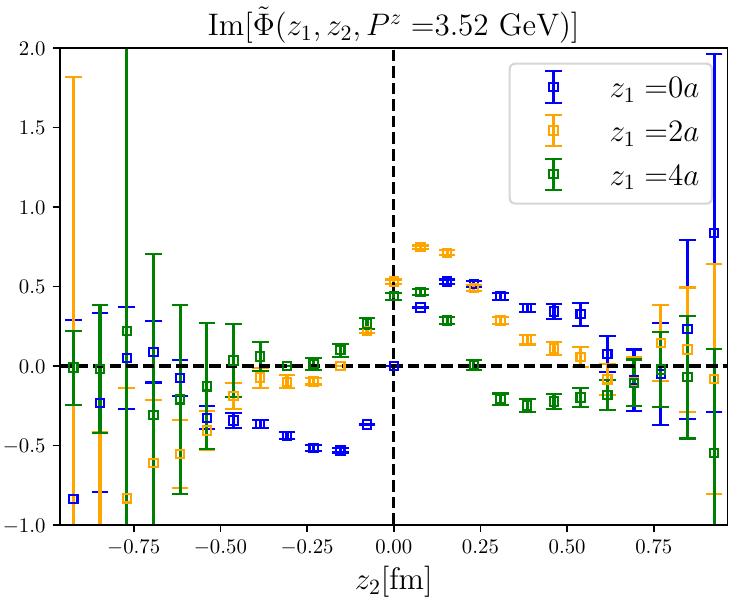}
\includegraphics[scale=0.63]{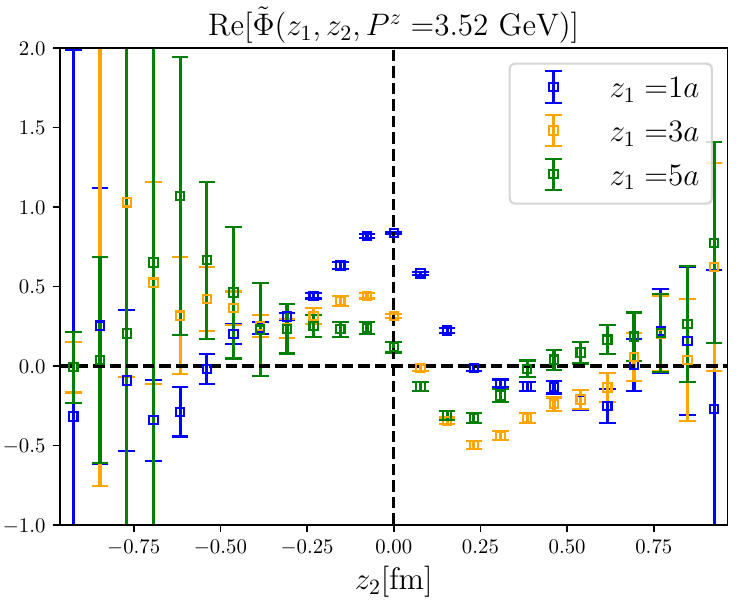}
\includegraphics[scale=0.63]{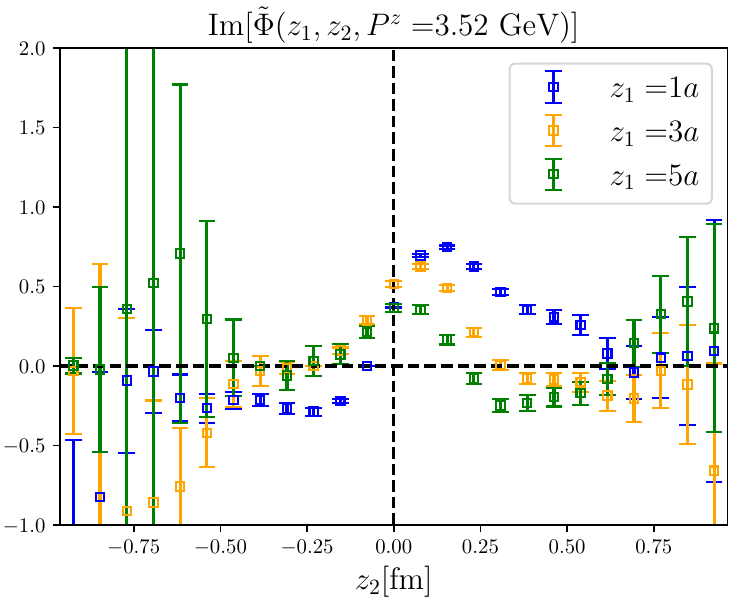}
\includegraphics[scale=0.63]{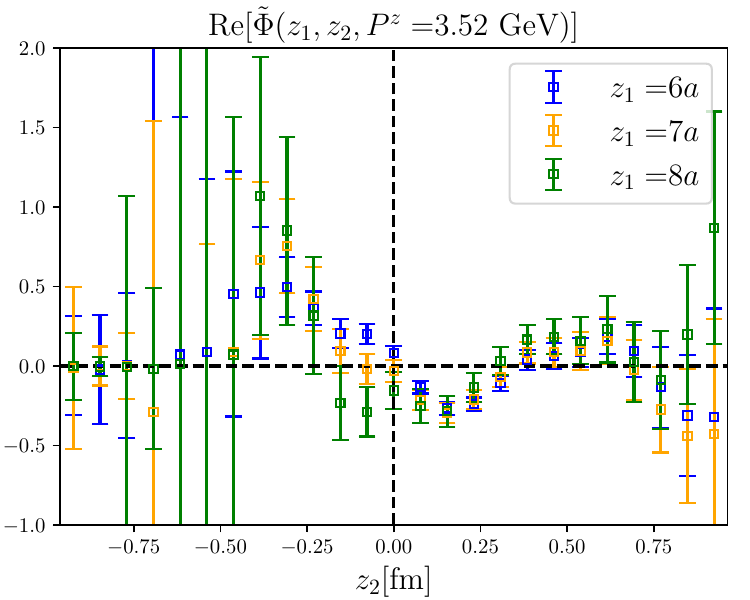}
\includegraphics[scale=0.63]{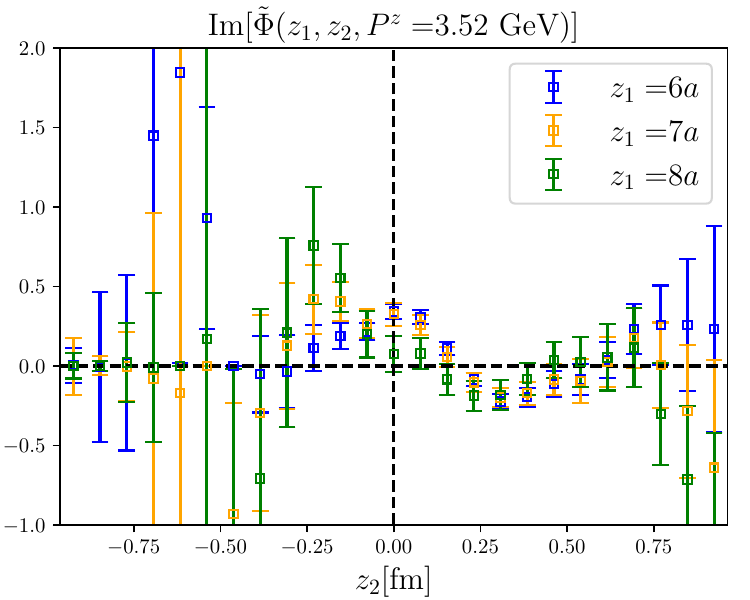}
\caption{The figures show the behavior of $\tilde{\Phi}(z_1,z_2)$ at all fixed $z_1\leq8a$ as a function of $z_2$. These are cases at $P^z=3.53$ GeV.}
\label{fig:qda_co_all_P7}
\end{figure}

In Sec.~\ref{sec:2pt_qda_co}, the behavior of the quasi-DA in coordinate space, $\tilde{\Phi}(z_1,z_2)$ is discussed. However, this discussion encompasses only a subset of all examples. As a supplement, we present the remaining cases here, including $P^z=\{2.58, 3.02, 3.52\}$ GeV for all $z_1$, illustrated in Figs.~\ref{fig:qda_co_all_P5}, \ref{fig:qda_co_all_P6}, and \ref{fig:qda_co_all_P7}. Due to significant uncertainties for  $z_1\geq 8a$ , this region does not yield relevant information. Therefore, only $z_1\leq 8a$ is displayed.

\end{widetext}


\clearpage

\begin{thebibliography}{}

\bibitem{Lepage:1980fj}
G.~P.~Lepage and S.~J.~Brodsky,
Phys. Rev. D \textbf{22}, 2157 (1980)
doi:10.1103/PhysRevD.22.2157

\bibitem{Chernyak:1983ej}
V.~L.~Chernyak and A.~R.~Zhitnitsky,
Phys. Rept. \textbf{112}, 173 (1984)
doi:10.1016/0370-1573(84)90126-1

\bibitem{LHCb:2016yco}
R.~Aaij \textit{et al.} [LHCb],
Nature Phys. \textbf{13}, 391-396 (2017)
doi:10.1038/nphys4021
[arXiv:1609.05216 [hep-ex]].

\bibitem{Bali:2017ude}
G.~S.~Bali \textit{et al.} [RQCD],
Phys. Lett. B \textbf{774}, 91-97 (2017)
doi:10.1016/j.physletb.2017.08.077
[arXiv:1705.10236 [hep-lat]].

\bibitem{RQCD:2019osh}
G.~S.~Bali \textit{et al.} [RQCD],
JHEP \textbf{08}, 065 (2019)
doi:10.1007/JHEP08(2019)065
[arXiv:1903.08038 [hep-lat]].

\bibitem{Zhang:2017bzy}
J.~H.~Zhang, J.~W.~Chen, X.~Ji, L.~Jin and H.~W.~Lin,
Phys. Rev. D \textbf{95}, no.9, 094514 (2017)
doi:10.1103/PhysRevD.95.094514
[arXiv:1702.00008 [hep-lat]].

\bibitem{Chen:2017gck}
J.~H.~Zhang \textit{et al.} [LP3],
Nucl. Phys. B \textbf{939}, 429-446 (2019)
doi:10.1016/j.nuclphysb.2018.12.020
[arXiv:1712.10025 [hep-ph]].

\bibitem{Zhang:2020gaj}
R.~Zhang, C.~Honkala, H.~W.~Lin and J.~W.~Chen,
Phys. Rev. D \textbf{102}, no.9, 094519 (2020)
doi:10.1103/PhysRevD.102.094519
[arXiv:2005.13955 [hep-lat]].

\bibitem{Gao:2022vyh}
X.~Gao, A.~D.~Hanlon, N.~Karthik, S.~Mukherjee, P.~Petreczky, P.~Scior, S.~Syritsyn and Y.~Zhao,
Phys. Rev. D \textbf{106}, no.7, 074505 (2022)
doi:10.1103/PhysRevD.106.074505
[arXiv:2206.04084 [hep-lat]].

\bibitem{Holligan:2023rex}
J.~Holligan, X.~Ji, H.~W.~Lin, Y.~Su and R.~Zhang,
Nucl. Phys. B \textbf{993}, 116282 (2023)
doi:10.1016/j.nuclphysb.2023.116282
[arXiv:2301.10372 [hep-lat]].

\bibitem{Hua:2020gnw}
J.~Hua \textit{et al.} [Lattice Parton],
Phys. Rev. Lett. \textbf{127}, no.6, 062002 (2021)
doi:10.1103/PhysRevLett.127.062002
[arXiv:2011.09788 [hep-lat]].

\bibitem{LatticeParton:2022zqc}
J.~Hua \textit{et al.} [Lattice Parton],
Phys. Rev. Lett. \textbf{129}, no.13, 132001 (2022)
doi:10.1103/PhysRevLett.129.132001
[arXiv:2201.09173 [hep-lat]].

\bibitem{Baker:2024zcd}
E.~Baker, D.~Bollweg, P.~Boyle, I.~Clo\"et, X.~Gao, S.~Mukherjee, P.~Petreczky, R.~Zhang and Y.~Zhao,
JHEP \textbf{07}, 211 (2024)
doi:10.1007/JHEP07(2024)211
[arXiv:2405.20120 [hep-lat]].

\bibitem{Cloet:2024vbv}
I.~Cloet, X.~Gao, S.~Mukherjee, S.~Syritsyn, N.~Karthik, P.~Petreczky, R.~Zhang and Y.~Zhao,
[arXiv:2407.00206 [hep-lat]].

\bibitem{Chernyak:1987nu}
V.~L.~Chernyak, A.~A.~Ogloblin and I.~R.~Zhitnitsky,
Yad. Fiz. \textbf{48}, 1410-1422 (1988)
doi:10.1007/BF01557663

\bibitem{Radyushkin:1990te}
A.~V.~Radyushkin,
Nucl. Phys. A \textbf{532}, 141-154 (1991)
doi:10.1016/0375-9474(91)90691-X

\bibitem{Bolz:1996sw}
J.~Bolz and P.~Kroll,
Z. Phys. A \textbf{356}, 327 (1996)
doi:10.1007/s002180050186
[arXiv:hep-ph/9603289 [hep-ph]].

\bibitem{Braun:2001tj}
V.~M.~Braun, A.~Lenz, N.~Mahnke and E.~Stein,
Phys. Rev. D \textbf{65}, 074011 (2002)
doi:10.1103/PhysRevD.65.074011
[arXiv:hep-ph/0112085 [hep-ph]].

\bibitem{Anikin:2013aka}
I.~V.~Anikin, V.~M.~Braun and N.~Offen,
Phys. Rev. D \textbf{88}, 114021 (2013)
doi:10.1103/PhysRevD.88.114021
[arXiv:1310.1375 [hep-ph]].

\bibitem{Bell:2013tfa}
G.~Bell, T.~Feldmann, Y.~M.~Wang and M.~W.~Y.~Yip,
JHEP \textbf{11} (2013), 191
doi:10.1007/JHEP11(2013)191
[arXiv:1308.6114 [hep-ph]].

\bibitem{Bali:2015ykx}
G.~S.~Bali, V.~M.~Braun, M.~G\"ockeler, M.~Gruber, F.~Hutzler, A.~Sch\"afer, R.~W.~Schiel, J.~Simeth, W.~S\"oldner and A.~Sternbeck, \textit{et al.}
JHEP \textbf{02}, 070 (2016)
doi:10.1007/JHEP02(2016)070
[arXiv:1512.02050 [hep-lat]].

\bibitem{RQCD:2019hps}
G.~S.~Bali \textit{et al.} [RQCD],
Eur. Phys. J. A \textbf{55}, no.7, 116 (2019)
doi:10.1140/epja/i2019-12803-6
[arXiv:1903.12590 [hep-lat]].

\bibitem{Chen:2024fhj}
W.~Chen, F.~Feng and Y.~Jia,
[arXiv:2406.19994 [hep-ph]].

\bibitem{Huang:2024ugd}
Y.~K.~Huang, B.~X.~Shi, Y.~M.~Wang and X.~C.~Zhao,
[arXiv:2407.18724 [hep-ph]].

\bibitem{Xiong:2013bka}
X.~Xiong, X.~Ji, J.~H.~Zhang and Y.~Zhao,
Phys. Rev. D \textbf{90}, no.1, 014051 (2014)
doi:10.1103/PhysRevD.90.014051
[arXiv:1310.7471 [hep-ph]].

\bibitem{Lin:2014zya}
H.~W.~Lin, J.~W.~Chen, S.~D.~Cohen and X.~Ji,
Phys. Rev. D \textbf{91}, 054510 (2015)
doi:10.1103/PhysRevD.91.054510
[arXiv:1402.1462 [hep-ph]].

\bibitem{Alexandrou:2015rja}
C.~Alexandrou, K.~Cichy, V.~Drach, E.~Garcia-Ramos, K.~Hadjiyiannakou, K.~Jansen, F.~Steffens and C.~Wiese,
Phys. Rev. D \textbf{92}, 014502 (2015)
doi:10.1103/PhysRevD.92.014502
[arXiv:1504.07455 [hep-lat]].

\bibitem{Chen:2016utp}
J.~W.~Chen, S.~D.~Cohen, X.~Ji, H.~W.~Lin and J.~H.~Zhang,
Nucl. Phys. B \textbf{911}, 246-273 (2016)
doi:10.1016/j.nuclphysb.2016.07.033
[arXiv:1603.06664 [hep-ph]].

\bibitem{Alexandrou:2016jqi}
C.~Alexandrou, K.~Cichy, M.~Constantinou, K.~Hadjiyiannakou, K.~Jansen, F.~Steffens and C.~Wiese,
Phys. Rev. D \textbf{96}, no.1, 014513 (2017)
doi:10.1103/PhysRevD.96.014513
[arXiv:1610.03689 [hep-lat]].

\bibitem{Alexandrou:2018pbm}
C.~Alexandrou, K.~Cichy, M.~Constantinou, K.~Jansen, A.~Scapellato and F.~Steffens,
Phys. Rev. Lett. \textbf{121}, no.11, 112001 (2018)
doi:10.1103/PhysRevLett.121.112001
[arXiv:1803.02685 [hep-lat]].

\bibitem{Chen:2018xof}
J.~W.~Chen, L.~Jin, H.~W.~Lin, Y.~S.~Liu, Y.~B.~Yang, J.~H.~Zhang and Y.~Zhao,
[arXiv:1803.04393 [hep-lat]].

\bibitem{Lin:2018pvv}
H.~W.~Lin, J.~W.~Chen, X.~Ji, L.~Jin, R.~Li, Y.~S.~Liu, Y.~B.~Yang, J.~H.~Zhang and Y.~Zhao,
Phys. Rev. Lett. \textbf{121}, no.24, 242003 (2018)
doi:10.1103/PhysRevLett.121.242003
[arXiv:1807.07431 [hep-lat]].

\bibitem{LatticeParton:2018gjr}
Y.~S.~Liu \textit{et al.} [Lattice Parton],
Phys. Rev. D \textbf{101}, no.3, 034020 (2020)
doi:10.1103/PhysRevD.101.034020
[arXiv:1807.06566 [hep-lat]].

\bibitem{Alexandrou:2018eet}
C.~Alexandrou, K.~Cichy, M.~Constantinou, K.~Jansen, A.~Scapellato and F.~Steffens,
Phys. Rev. D \textbf{98}, no.9, 091503 (2018)
doi:10.1103/PhysRevD.98.091503
[arXiv:1807.00232 [hep-lat]].

\bibitem{Liu:2018hxv}
Y.~S.~Liu, J.~W.~Chen, L.~Jin, R.~Li, H.~W.~Lin, Y.~B.~Yang, J.~H.~Zhang and Y.~Zhao,
[arXiv:1810.05043 [hep-lat]].

\bibitem{Chen:2018fwa}
J.~H.~Zhang, J.~W.~Chen, L.~Jin, H.~W.~Lin, A.~Sch\"afer and Y.~Zhao,
Phys. Rev. D \textbf{100}, no.3, 034505 (2019)
doi:10.1103/PhysRevD.100.034505
[arXiv:1804.01483 [hep-lat]].

\bibitem{Izubuchi:2018srq}
T.~Izubuchi, X.~Ji, L.~Jin, I.~W.~Stewart and Y.~Zhao,
Phys. Rev. D \textbf{98}, no.5, 056004 (2018)
doi:10.1103/PhysRevD.98.056004
[arXiv:1801.03917 [hep-ph]].

\bibitem{Izubuchi:2019lyk}
T.~Izubuchi, L.~Jin, C.~Kallidonis, N.~Karthik, S.~Mukherjee, P.~Petreczky, C.~Shugert and S.~Syritsyn,
Phys. Rev. D \textbf{100}, no.3, 034516 (2019)
doi:10.1103/PhysRevD.100.034516
[arXiv:1905.06349 [hep-lat]].

\bibitem{Shugert:2020tgq}
C.~Shugert, X.~Gao, T.~Izubichi, L.~Jin, C.~Kallidonis, N.~Karthik, S.~Mukherjee, P.~Petreczky, S.~Syritsyn and Y.~Zhao,
[arXiv:2001.11650 [hep-lat]].

\bibitem{Chai:2020nxw}
Y.~Chai, Y.~Li, S.~Xia, C.~Alexandrou, K.~Cichy, M.~Constantinou, X.~Feng, K.~Hadjiyiannakou, K.~Jansen and G.~Koutsou, \textit{et al.}
Phys. Rev. D \textbf{102}, no.1, 014508 (2020)
doi:10.1103/PhysRevD.102.014508
[arXiv:2002.12044 [hep-lat]].

\bibitem{Lin:2020ssv}
H.~W.~Lin, J.~W.~Chen, Z.~Fan, J.~H.~Zhang and R.~Zhang,
Phys. Rev. D \textbf{103}, no.1, 014516 (2021)
doi:10.1103/PhysRevD.103.014516
[arXiv:2003.14128 [hep-lat]].

\bibitem{Fan:2020nzz}
Z.~Fan, X.~Gao, R.~Li, H.~W.~Lin, N.~Karthik, S.~Mukherjee, P.~Petreczky, S.~Syritsyn, Y.~B.~Yang and R.~Zhang,
Phys. Rev. D \textbf{102}, no.7, 074504 (2020)
doi:10.1103/PhysRevD.102.074504
[arXiv:2005.12015 [hep-lat]].

\bibitem{Gao:2021hxl}
X.~Gao, K.~Lee, S.~Mukherjee, C.~Shugert and Y.~Zhao,
Phys. Rev. D \textbf{103}, no.9, 094504 (2021)
doi:10.1103/PhysRevD.103.094504
[arXiv:2102.01101 [hep-ph]].

\bibitem{Gao:2021dbh}
X.~Gao, A.~D.~Hanlon, S.~Mukherjee, P.~Petreczky, P.~Scior, S.~Syritsyn and Y.~Zhao,
Phys. Rev. Lett. \textbf{128}, no.14, 142003 (2022)
doi:10.1103/PhysRevLett.128.142003
[arXiv:2112.02208 [hep-lat]].

\bibitem{Gao:2022iex}
X.~Gao, A.~D.~Hanlon, N.~Karthik, S.~Mukherjee, P.~Petreczky, P.~Scior, S.~Shi, S.~Syritsyn, Y.~Zhao and K.~Zhou,
Phys. Rev. D \textbf{106}, no.11, 114510 (2022)
doi:10.1103/PhysRevD.106.114510
[arXiv:2208.02297 [hep-lat]].

\bibitem{Su:2022fiu}
Y.~Su, J.~Holligan, X.~Ji, F.~Yao, J.~H.~Zhang and R.~Zhang,
Nucl. Phys. B \textbf{991}, 116201 (2023)
doi:10.1016/j.nuclphysb.2023.116201
[arXiv:2209.01236 [hep-ph]].

\bibitem{LatticeParton:2022xsd}
F.~Yao \textit{et al.} [Lattice Parton],
Phys. Rev. Lett. \textbf{131}, no.26, 261901 (2023)
doi:10.1103/PhysRevLett.131.261901
[arXiv:2208.08008 [hep-lat]].

\bibitem{Gao:2022uhg}
X.~Gao, A.~D.~Hanlon, J.~Holligan, N.~Karthik, S.~Mukherjee, P.~Petreczky, S.~Syritsyn and Y.~Zhao,
Phys. Rev. D \textbf{107}, no.7, 074509 (2023)
doi:10.1103/PhysRevD.107.074509
[arXiv:2212.12569 [hep-lat]].

\bibitem{Chou:2022drv}
C.~Y.~Chou and J.~W.~Chen,
Phys. Rev. D \textbf{106}, no.1, 014507 (2022)
doi:10.1103/PhysRevD.106.014507
[arXiv:2204.08343 [hep-lat]].

\bibitem{Gao:2023lny}
X.~Gao, W.~Y.~Liu and Y.~Zhao,
Phys. Rev. D \textbf{109}, no.9, 094506 (2024)
doi:10.1103/PhysRevD.109.094506
[arXiv:2306.14960 [hep-ph]].

\bibitem{Gao:2023ktu}
X.~Gao, A.~D.~Hanlon, S.~Mukherjee, P.~Petreczky, Q.~Shi, S.~Syritsyn and Y.~Zhao,
Phys. Rev. D \textbf{109}, no.5, 054506 (2024)
doi:10.1103/PhysRevD.109.054506
[arXiv:2310.19047 [hep-lat]].

\bibitem{Chen:2024rgi}
C.~Chen, L.~Liu, P.~Sun, Y.~B.~Yang, Y.~Geng, F.~Yao, J.~H.~Zhang and K.~Zhang,
[arXiv:2408.12819 [hep-lat]].

\bibitem{Holligan:2024umc}
J.~Holligan and H.~W.~Lin,
J. Phys. G \textbf{51}, no.6, 065101 (2024)
doi:10.1088/1361-6471/ad3162
[arXiv:2404.14525 [hep-lat]].

\bibitem{Holligan:2024wpv}
J.~Holligan and H.~W.~Lin,
Phys. Lett. B \textbf{854}, 138731 (2024)
doi:10.1016/j.physletb.2024.138731
[arXiv:2405.18238 [hep-lat]].

\bibitem{Wang:2017qyg}
W.~Wang, S.~Zhao and R.~Zhu,
Eur. Phys. J. C \textbf{78}, no.2, 147 (2018)
doi:10.1140/epjc/s10052-018-5617-3
[arXiv:1708.02458 [hep-ph]].

\bibitem{Wang:2017eel}
W.~Wang and S.~Zhao,
JHEP \textbf{05}, 142 (2018)
doi:10.1007/JHEP05(2018)142
[arXiv:1712.09247 [hep-ph]].

\bibitem{Fan:2018dxu}
Z.~Y.~Fan, Y.~B.~Yang, A.~Anthony, H.~W.~Lin and K.~F.~Liu,
Phys. Rev. Lett. \textbf{121}, no.24, 242001 (2018)
doi:10.1103/PhysRevLett.121.242001
[arXiv:1808.02077 [hep-lat]].

\bibitem{Wang:2019tgg}
W.~Wang, J.~H.~Zhang, S.~Zhao and R.~Zhu,
Phys. Rev. D \textbf{100}, no.7, 074509 (2019)
doi:10.1103/PhysRevD.100.074509
[arXiv:1904.00978 [hep-ph]].

\bibitem{Good:2024iur}
W.~Good, K.~Hasan and H.~W.~Lin,
[arXiv:2409.02750 [hep-lat]].

\bibitem{Chen:2019lcm}
J.~W.~Chen, H.~W.~Lin and J.~H.~Zhang,
Nucl. Phys. B \textbf{952}, 114940 (2020)
doi:10.1016/j.nuclphysb.2020.114940
[arXiv:1904.12376 [hep-lat]].

\bibitem{Alexandrou:2019dax}
C.~Alexandrou, K.~Cichy, M.~Constantinou, K.~Hadjiyiannakou, K.~Jansen, A.~Scapellato and F.~Steffens,
PoS \textbf{LATTICE2019}, 036 (2019)
doi:10.22323/1.363.0036
[arXiv:1910.13229 [hep-lat]].

\bibitem{Lin:2020rxa}
H.~W.~Lin,
Phys. Rev. Lett. \textbf{127}, no.18, 182001 (2021)
doi:10.1103/PhysRevLett.127.182001
[arXiv:2008.12474 [hep-ph]].

\bibitem{Alexandrou:2020zbe}
C.~Alexandrou, K.~Cichy, M.~Constantinou, K.~Hadjiyiannakou, K.~Jansen, A.~Scapellato and F.~Steffens,
Phys. Rev. Lett. \textbf{125}, no.26, 262001 (2020)
doi:10.1103/PhysRevLett.125.262001
[arXiv:2008.10573 [hep-lat]].

\bibitem{Lin:2021brq}
H.~W.~Lin,
Phys. Lett. B \textbf{824}, 136821 (2022)
doi:10.1016/j.physletb.2021.136821
[arXiv:2112.07519 [hep-lat]].

\bibitem{Scapellato:2022mai}
A.~Scapellato, C.~Alexandrou, K.~Cichy, M.~Constantinou, K.~Hadjiyiannakou, K.~Jansen and F.~Steffens,
Rev. Mex. Fis. Suppl. \textbf{3}, no.3, 0308104 (2022)
doi:10.31349/SuplRevMexFis.3.0308104
[arXiv:2201.06519 [hep-lat]].

\bibitem{Bhattacharya:2022aob}
S.~Bhattacharya, K.~Cichy, M.~Constantinou, J.~Dodson, X.~Gao, A.~Metz, S.~Mukherjee, A.~Scapellato, F.~Steffens and Y.~Zhao,
Phys. Rev. D \textbf{106}, no.11, 114512 (2022)
doi:10.1103/PhysRevD.106.114512
[arXiv:2209.05373 [hep-lat]].

\bibitem{Bhattacharya:2023nmv}
S.~Bhattacharya, K.~Cichy, M.~Constantinou, J.~Dodson, A.~Metz, A.~Scapellato and F.~Steffens,
Phys. Rev. D \textbf{108}, no.5, 054501 (2023)
doi:10.1103/PhysRevD.108.054501
[arXiv:2306.05533 [hep-lat]].

\bibitem{Bhattacharya:2023jsc}
S.~Bhattacharya, K.~Cichy, M.~Constantinou, J.~Dodson, X.~Gao, A.~Metz, J.~Miller, S.~Mukherjee, P.~Petreczky and F.~Steffens, \textit{et al.}
Phys. Rev. D \textbf{109}, no.3, 034508 (2024)
doi:10.1103/PhysRevD.109.034508
[arXiv:2310.13114 [hep-lat]].

\bibitem{Lin:2023gxz}
H.~W.~Lin,
Phys. Lett. B \textbf{846}, 138181 (2023)
doi:10.1016/j.physletb.2023.138181
[arXiv:2310.10579 [hep-lat]].

\bibitem{Holligan:2023jqh}
J.~Holligan and H.~W.~Lin,
Phys. Rev. D \textbf{110}, no.3, 3 (2024)
doi:10.1103/PhysRevD.110.034503
[arXiv:2312.10829 [hep-lat]].

\bibitem{Ding:2024hkz}
H.~T.~Ding, X.~Gao, S.~Mukherjee, P.~Petreczky, Q.~Shi, S.~Syritsyn and Y.~Zhao,
[arXiv:2407.03516 [hep-lat]].

\bibitem{Liu:2018tox}
Y.~S.~Liu, W.~Wang, J.~Xu, Q.~A.~Zhang, S.~Zhao and Y.~Zhao,
Phys. Rev. D \textbf{99}, no.9, 094036 (2019)
doi:10.1103/PhysRevD.99.094036
[arXiv:1810.10879 [hep-ph]].

\bibitem{Deng:2023csv}
Z.~F.~Deng, C.~Han, W.~Wang, J.~Zeng and J.~L.~Zhang,
JHEP \textbf{07}, 191 (2023)
doi:10.1007/JHEP07(2023)191
[arXiv:2304.09004 [hep-ph]].

\bibitem{Han:2023xbl}
C.~Han, Y.~Su, W.~Wang and J.~L.~Zhang,
JHEP \textbf{12}, 044 (2023)
doi:10.1007/JHEP12(2023)044
[arXiv:2308.16793 [hep-ph]].

\bibitem{Han:2023hgy}
C.~Han and J.~Zhang,
Phys. Rev. D \textbf{109}, no.1, 014034 (2024)
doi:10.1103/PhysRevD.109.014034
[arXiv:2311.02669 [hep-ph]].

\bibitem{Han:2024ucv}
C.~Han, W.~Wang, J.~Zeng and J.~L.~Zhang,
JHEP \textbf{07}, 019 (2024)
doi:10.1007/JHEP07(2024)019
[arXiv:2404.04855 [hep-ph]].

\bibitem{Wang:2019msf}
W.~Wang, Y.~M.~Wang, J.~Xu and S.~Zhao,
Phys. Rev. D \textbf{102}, no.1, 011502 (2020)
doi:10.1103/PhysRevD.102.011502
[arXiv:1908.09933 [hep-ph]].

\bibitem{Zhao:2020bsx}
S.~Zhao and A.~V.~Radyushkin,
Phys. Rev. D \textbf{103}, no.5, 054022 (2021)
doi:10.1103/PhysRevD.103.054022
[arXiv:2006.05663 [hep-ph]].

\bibitem{Xu:2022krn}
J.~Xu, X.~R.~Zhang and S.~Zhao,
Phys. Rev. D \textbf{106}, no.1, L011503 (2022)
doi:10.1103/PhysRevD.106.L011503
[arXiv:2202.13648 [hep-ph]].

\bibitem{Xu:2022guw}
J.~Xu and X.~R.~Zhang,
Phys. Rev. D \textbf{106}, no.11, 114019 (2022)
doi:10.1103/PhysRevD.106.114019
[arXiv:2209.10719 [hep-ph]].

\bibitem{Hu:2023bba}
S.~M.~Hu, W.~Wang, J.~Xu and S.~Zhao,
Phys. Rev. D \textbf{109}, no.3, 034001 (2024)
doi:10.1103/PhysRevD.109.034001
[arXiv:2308.13977 [hep-ph]].

\bibitem{Hu:2024ebp}
S.~M.~Hu, J.~Xu and S.~Zhao,
Eur. Phys. J. C \textbf{84}, no.5, 502 (2024)
doi:10.1140/epjc/s10052-024-12672-2
[arXiv:2401.04291 [hep-ph]].

\bibitem{Han:2024min}
X.~Y.~Han, J.~Hua, X.~Ji, C.~D.~L\"u, W.~Wang, J.~Xu, Q.~A.~Zhang and S.~Zhao,
[arXiv:2403.17492 [hep-ph]].

\bibitem{Han:2024cht}
C.~Han, W.~Wang, J.~L.~Zhang and J.~H.~Zhang,
[arXiv:2408.13486 [hep-ph]].

\bibitem{Deng:2024dkd}
Z.~F.~Deng, W.~Wang, Y.~B.~Wei and J.~Zeng,
[arXiv:2409.00632 [hep-ph]].

\bibitem{Han:2024yun}
X.~Y.~Han, J.~Hua, X.~Ji, C.~D.~L\"u, A.~Sch\"afer, Y.~Su, W.~Wang, J.~Xu, Y.~Yang and J.~H.~Zhang, \textit{et al.}
[arXiv:2410.18654 [hep-lat]].

\bibitem{Ji:2014hxa}
X.~Ji, P.~Sun, X.~Xiong and F.~Yuan,
Phys. Rev. D \textbf{91}, 074009 (2015)
doi:10.1103/PhysRevD.91.074009
[arXiv:1405.7640 [hep-ph]].

\bibitem{Shanahan:2019zcq}
P.~Shanahan, M.~L.~Wagman and Y.~Zhao,
Phys. Rev. D \textbf{101}, no.7, 074505 (2020)
doi:10.1103/PhysRevD.101.074505
[arXiv:1911.00800 [hep-lat]].

\bibitem{Shanahan:2020zxr}
P.~Shanahan, M.~Wagman and Y.~Zhao,
Phys. Rev. D \textbf{102}, no.1, 014511 (2020)
doi:10.1103/PhysRevD.102.014511
[arXiv:2003.06063 [hep-lat]].

\bibitem{Zhang:2020dbb}
Q.~A.~Zhang \textit{et al.} [Lattice Parton],
Phys. Rev. Lett. \textbf{125}, no.19, 192001 (2020)
doi:10.22323/1.396.0477
[arXiv:2005.14572 [hep-lat]].

\bibitem{Ji:2021znw}
X.~Ji and Y.~Liu,
Phys. Rev. D \textbf{105}, no.7, 076014 (2022)
doi:10.1103/PhysRevD.105.076014
[arXiv:2106.05310 [hep-ph]].

\bibitem{LatticePartonLPC:2022eev}
M.~H.~Chu \textit{et al.} [Lattice Parton (LPC)],
Phys. Rev. D \textbf{106}, no.3, 034509 (2022)
doi:10.1103/PhysRevD.106.034509
[arXiv:2204.00200 [hep-lat]].

\bibitem{Liu:2022nnk}
Y.~Liu,
Acta Phys. Polon. B \textbf{53}, no.4, 4-A2 (2022)
doi:10.5506/APhysPolB.53.4-A2

\bibitem{Zhang:2022xuw}
K.~Zhang \textit{et al.} [Lattice Parton (LPC)],
Phys. Rev. Lett. \textbf{129}, no.8, 082002 (2022)
doi:10.1103/PhysRevLett.129.082002
[arXiv:2205.13402 [hep-lat]].

\bibitem{Deng:2022gzi}
Z.~F.~Deng, W.~Wang and J.~Zeng,
JHEP \textbf{09}, 046 (2022)
doi:10.1007/JHEP09(2022)046
[arXiv:2207.07280 [hep-th]].

\bibitem{Zhu:2022bja}
R.~Zhu, Y.~Ji, J.~H.~Zhang and S.~Zhao,
JHEP \textbf{02}, 114 (2023)
doi:10.1007/JHEP02(2023)114
[arXiv:2209.05443 [hep-ph]].

\bibitem{LatticePartonCollaborationLPC:2022myp}
J.~C.~He \textit{et al.} [Lattice Parton Collaboration (LPC)],
Phys. Rev. D \textbf{109}, no.11, 114513 (2024)
doi:10.1103/PhysRevD.109.114513
[arXiv:2211.02340 [hep-lat]].

\bibitem{Rodini:2022wic}
S.~Rodini and A.~Vladimirov,
JHEP \textbf{09}, 117 (2023)
doi:10.1007/JHEP09(2023)117
[arXiv:2211.04494 [hep-ph]].

\bibitem{Shu:2023cot}
H.~T.~Shu, M.~Schlemmer, T.~Sizmann, A.~Vladimirov, L.~Walter, M.~Engelhardt, A.~Sch\"afer and Y.~B.~Yang,
Phys. Rev. D \textbf{108}, no.7, 074519 (2023)
doi:10.1103/PhysRevD.108.074519
[arXiv:2302.06502 [hep-lat]].

\bibitem{Chu:2023jia}
M.~H.~Chu \textit{et al.} [Lattice Parton],
Phys. Rev. D \textbf{109}, no.9, L091503 (2024)
doi:10.1103/PhysRevD.109.L091503
[arXiv:2302.09961 [hep-lat]].

\bibitem{delRio:2023pse}
\'O.~del R\'\i{}o and A.~Vladimirov,
Phys. Rev. D \textbf{108}, no.11, 114009 (2023)
doi:10.1103/PhysRevD.108.114009
[arXiv:2304.14440 [hep-ph]].

\bibitem{LatticePartonLPC:2023pdv}
M.~H.~Chu \textit{et al.} [Lattice Parton (LPC)],
JHEP \textbf{08}, 172 (2023)
doi:10.1007/JHEP08(2023)172
[arXiv:2306.06488 [hep-lat]].

\bibitem{LatticeParton:2023xdl}
M.~H.~Chu \textit{et al.} [Lattice Parton],
Phys. Rev. D \textbf{109}, no.9, L091503 (2024)
doi:10.1103/PhysRevD.109.L091503
[arXiv:2302.09961 [hep-lat]].

\bibitem{Alexandrou:2023ucc}
C.~Alexandrou, S.~Bacchio, K.~Cichy, M.~Constantinou, X.~Feng, K.~Jansen, C.~Liu, A.~Sen, G.~Spanoudes and F.~Steffens, \textit{et al.}
Phys. Rev. D \textbf{108}, no.11, 114503 (2023)
doi:10.1103/PhysRevD.108.114503
[arXiv:2305.11824 [hep-lat]].

\bibitem{Avkhadiev:2023poz}
A.~Avkhadiev, P.~E.~Shanahan, M.~L.~Wagman and Y.~Zhao,
Phys. Rev. D \textbf{108}, no.11, 114505 (2023)
doi:10.1103/PhysRevD.108.114505
[arXiv:2307.12359 [hep-lat]].

\bibitem{Zhao:2023ptv}
Y.~Zhao,
[arXiv:2311.01391 [hep-ph]].

\bibitem{Avkhadiev:2024mgd}
A.~Avkhadiev, P.~E.~Shanahan, M.~L.~Wagman and Y.~Zhao,
Phys. Rev. Lett. \textbf{132}, no.23, 231901 (2024)
doi:10.1103/PhysRevLett.132.231901
[arXiv:2402.06725 [hep-lat]].

\bibitem{Bollweg:2024zet}
D.~Bollweg, X.~Gao, S.~Mukherjee and Y.~Zhao,
Phys. Lett. B \textbf{852}, 138617 (2024)
doi:10.1016/j.physletb.2024.138617
[arXiv:2403.00664 [hep-lat]].

\bibitem{Spanoudes:2024kpb}
G.~Spanoudes, M.~Constantinou and H.~Panagopoulos,
Phys. Rev. D \textbf{109}, no.11, 114501 (2024)
doi:10.1103/PhysRevD.109.114501
[arXiv:2401.01182 [hep-lat]].

\bibitem{Zhang:2023wea}
J.~H.~Zhang,
[arXiv:2304.12481 [hep-ph]].

\bibitem{Jaarsma:2023woo}
M.~Jaarsma, R.~Rahn and W.~J.~Waalewijn,
JHEP \textbf{12}, 014 (2023)
doi:10.1007/JHEP12(2023)014
[arXiv:2305.09716 [hep-ph]].

\bibitem{Braun:2000kw}
V.~Braun, R.~J.~Fries, N.~Mahnke and E.~Stein,
Nucl. Phys. B \textbf{589}, 381-409 (2000)
[erratum: Nucl. Phys. B \textbf{607}, 433-433 (2001)]
doi:10.1016/S0550-3213(00)00516-2
[arXiv:hep-ph/0007279 [hep-ph]].

\bibitem{Zhang:2017zfe}
J.~H.~Zhang \textit{et al.} [LP3],
Nucl. Phys. B \textbf{939}, 429-446 (2019)
doi:10.1016/j.nuclphysb.2018.12.020
[arXiv:1712.10025 [hep-ph]].

\bibitem{Martinelli:1994ty}
G.~Martinelli, C.~Pittori, C.~T.~Sachrajda, M.~Testa and A.~Vladikas,
Nucl. Phys. B \textbf{445}, 81-108 (1995)
doi:10.1016/0550-3213(95)00126-D
[arXiv:hep-lat/9411010 [hep-lat]].

\bibitem{Ji:2020brr}
X.~Ji, Y.~Liu, A.~Sch\"afer, W.~Wang, Y.~B.~Yang, J.~H.~Zhang and Y.~Zhao,
Nucl. Phys. B \textbf{964}, 115311 (2021)
doi:10.1016/j.nuclphysb.2021.115311
[arXiv:2008.03886 [hep-ph]].

\bibitem{LatticePartonLPC:2021gpi}
Y.~K.~Huo \textit{et al.} [Lattice Parton (LPC)],
Nucl. Phys. B \textbf{969}, 115443 (2021)
doi:10.1016/j.nuclphysb.2021.115443
[arXiv:2103.02965 [hep-lat]].

\bibitem{Zhang:2024omt}
K.~Zhang \textit{et al.} [Lattice Parton],
Phys. Rev. D \textbf{110}, no.7, 074505 (2024)
doi:10.1103/PhysRevD.110.074505
[arXiv:2405.14097 [hep-lat]].

\bibitem{CLQCD:2023sdb}
Z.~C.~Hu \textit{et al.} [CLQCD],
Phys. Rev. D \textbf{109}, no.5, 054507 (2024)
doi:10.1103/PhysRevD.109.054507
[arXiv:2310.00814 [hep-lat]].

\bibitem{Liu:2022gxf}
H.~Liu, J.~He, L.~Liu, P.~Sun, W.~Wang, Y.~B.~Yang and Q.~A.~Zhang,
Sci. China Phys. Mech. Astron. \textbf{67}, no.1, 211011 (2024)
doi:10.1007/s11433-023-2205-0
[arXiv:2207.00183 [hep-lat]].

\bibitem{Xing:2022ijm}
H.~Xing, J.~Liang, L.~Liu, P.~Sun and Y.~B.~Yang,
[arXiv:2210.08555 [hep-lat]].

\bibitem{Yan:2024yuq}
H.~Yan, C.~Liu, L.~Liu, Y.~Meng and H.~Xing,
[arXiv:2404.13479 [hep-lat]].

\bibitem{Zhang:2021oja}
Q.~A.~Zhang, J.~Hua, F.~Huang, R.~Li, Y.~Li, C.~L\"u, C.~D.~Lu, P.~Sun, W.~Sun and W.~Wang, \textit{et al.}
Chin. Phys. C \textbf{46}, no.1, 011002 (2022)
doi:10.1088/1674-1137/ac2b12
[arXiv:2103.07064 [hep-lat]].

\bibitem{Liu:2023feb}
H.~Liu, L.~Liu, P.~Sun, W.~Sun, J.~X.~Tan, W.~Wang, Y.~B.~Yang and Q.~A.~Zhang,
Phys. Lett. B \textbf{841}, 137941 (2023)
doi:10.1016/j.physletb.2023.137941
[arXiv:2303.17865 [hep-lat]].

\bibitem{Liu:2023pwr}
H.~Liu, W.~Wang and Q.~A.~Zhang,
Phys. Rev. D \textbf{109}, no.3, 036037 (2024)
doi:10.1103/PhysRevD.109.036037
[arXiv:2309.05432 [hep-ph]].

\bibitem{Meng:2024nyo}
Y.~Meng, J.~L.~Dang, C.~Liu, X.~Y.~Tuo, H.~Yan, Y.~B.~Yang and K.~L.~Zhang,
Phys. Rev. D \textbf{110}, no.7, 074510 (2024)
doi:10.1103/PhysRevD.110.074510
[arXiv:2407.13568 [hep-lat]].

\bibitem{Du:2024wtr}
H.~Y.~Du \textit{et al.} [CLQCD],
[arXiv:2408.03548 [hep-lat]].

\bibitem{Zhao:2022ooq}
D.~J.~Zhao \textit{et al.} [\ensuremath{\chi}QCD],
Phys. Rev. D \textbf{107}, no.9, L091501 (2023)
doi:10.1103/PhysRevD.107.L091501
[arXiv:2207.14132 [hep-lat]].

\bibitem{Meng:2023nxf}
X.~L.~Meng \textit{et al.} [\ensuremath{\chi}QCD and CLQCD],
[arXiv:2305.09459 [hep-lat]].

\bibitem{Bali:2016lva}
G.~S.~Bali, B.~Lang, B.~U.~Musch and A.~Sch\"afer,
Phys. Rev. D \textbf{93} (2016) no.9, 094515
doi:10.1103/PhysRevD.93.094515
[arXiv:1602.05525 [hep-lat]].

\bibitem{APE:1987ehd}
M.~Albanese \textit{et al.} [APE],
Phys. Lett. B \textbf{192} (1987), 163-169
doi:10.1016/0370-2693(87)91160-9

\bibitem{Wang:2008sm}
Y.~m.~Wang, Y.~Li and C.~D.~Lu,
Eur. Phys. J. C \textbf{59}, 861-882 (2009)
doi:10.1140/epjc/s10052-008-0846-5
[arXiv:0804.0648 [hep-ph]].

\bibitem{LHCb:2015tgy}
R.~Aaij \textit{et al.} [LHCb],
JHEP \textbf{06}, 115 (2015)
[erratum: JHEP \textbf{09}, 145 (2018)]
doi:10.1007/JHEP06(2015)115
[arXiv:1503.07138 [hep-ex]].

\bibitem{Edwards:2004sx}
R.~G.~Edwards \textit{et al.} [SciDAC, LHPC and UKQCD],
Nucl. Phys. B Proc. Suppl. \textbf{140}, 832 (2005)
doi:10.1016/j.nuclphysbps.2004.11.254
[arXiv:hep-lat/0409003 [hep-lat]].

\bibitem{Clark:2009wm}
M.~A.~Clark \textit{et al.} [QUDA],
Comput. Phys. Commun. \textbf{181}, 1517-1528 (2010)
doi:10.1016/j.cpc.2010.05.002
[arXiv:0911.3191 [hep-lat]].

\bibitem{Babich:2011np}
R.~Babich \textit{et al.} [QUDA],
doi:10.1145/2063384.2063478
[arXiv:1109.2935 [hep-lat]].

\bibitem{Clark:2016rdz}
M.~A.~Clark \textit{et al.} [QUDA],
doi:10.5555/3014904.3014995
[arXiv:1612.07873 [hep-lat]].

\bibitem{Bi:2020wpt}
Y.~J.~Bi, Y.~Xiao, W.~Y.~Guo, M.~Gong, P.~Sun, S.~Xu and Y.~B.~Yang,
PoS \textbf{LATTICE2019}, 286 (2020)
doi:10.22323/1.363.0286
[arXiv:2001.05706 [hep-lat]].

\end{thebibliography}
\end{document}